# Independent Approximates
## enable closed-form estimation
## of heavy-tailed distributions

Kenric P. Nelson[1]

*Abstract – A new statistical estimation method, Independent Approximates (IAs), is defined and proven to enable closed-form estimation of the parameters of heavy-tailed distributions. Given independent, identically distributed samples from a one-dimensional distribution, IAs are formed by partitioning samples into pairs, triplets, or $n^{th}$-order groupings and retaining the median of those groupings that are approximately equal. The pdf of the IAs is proven to be the normalized $n^{th}$ power of the original density. From this property, heavy-tailed distributions are proven to have well-defined means for their IA pairs, finite second moments for their IA triplets, and a finite, well-defined $(n-1)^{th}$ moment for the $n^{th}$ grouping. Estimation of the location, scale, and shape (inverse of degree of freedom) of the generalized Pareto and Student's t distributions are possible via a system of three equations. Performance analysis of the IA estimation methodology for the Student's t distribution demonstrates that the method converges to the maximum likelihood estimate. Closed-form estimates of the location and scale are determined from the mean of the IA pairs and the second moment of the IA triplets, respectively. For the Student's t distribution, the geometric mean of the original samples provides a third equation to determine the shape, though its nonlinear solution requires an iterative solver. With 10,000 samples the relative bias of the parameter estimates is less than 0.01 and the relative precision is less than ±0.1. Statistical physics applications are carried out for both a small sample (331) astrophysics dataset and a large sample ($2 \times 10^8$) standard map simulation.*

**MCS**: 62F10

## Statements and Declarations

**Conflicts of interest:** Kenric Nelson declares there is no conflict of interest.
**Animal/Human Rights:** This article does not contain any studies with human or animal subjects performed by the author.
**Availability of code, data and material:**
https://github.com/Photrek/Independent-Approximates

---

[1] Kenric P. Nelson, Photrek, Watertown, MA USA 02472, kenric.nelson@photrek.world

## 1. Introduction

Statistical estimation of the parameters of a heavy-tailed distribution is an infamously difficult problem. As the rate of tail decay decreases, i.e. becomes heavier, the probability of outlier samples, which overwhelm statistical analysis, increases. As a result, the statistical moments either diverge to infinity or become undefined with the higher-order moments becoming unstable first. Nevertheless, the study of complex adaptive systems has shown that the nonlinear dynamics of both natural and man-made systems make heavy-tailed distributions ubiquitous and thus essential for accurate modeling of important signals and systems (Clauset et al. 2009). A few examples include the distribution of ecological systems (generalized Pareto) (Katz et al. 2005), the foraging patterns on birds (Levy) (Viswanathan et al. 1996), log-return of markets (Student's $t$) (Pisarenko and Sornette 2006; Vilela et al. 2019), and the distribution of words in a language (Zipf's Law) (Piantadosi 2014). Comprehensive surveys of statistical analysis of heavy-tailed distributions include Resnick (Resnick 2007), Kotz and Nadarajah (Kotz and Nadarajah 2004) and of modeling for complex systems include Aschwanden (Aschwanden 2011), Tsallis (Tsallis 2009), and Cirillo (Cirillo 2012). Despite the extensive investigation of heavy-tailed distributions and the importance of modeling the stochastic properties of nonlinear systems, estimation using closed-form solutions as opposed to iterative optimization has been limited.

Investigators studying nonextensive statistical mechanics (Tsallis 2009) and information geometry (Amari et al. 2012) have shown the importance of an escort density ${f^q(x)}/{\int f^q(x)}$ in characterizing a broad family of heavy-tailed distributions. While the parameter $q$ is referred to as the entropic index (Beck 2009), based on its generalization of thermodynamics and information theory; is related to fluctuations in the variance (Wilk and Włodarczyk 2000); and triplets of the parameter have been used to characterize complex systems (Tsallis 2006), none of these interpretations specify what physical property the parameter $q$ quantifies. Furthermore, while the escort density leads to a set of generalized moments, estimators based on the escort density must iterate between estimating the generalized moments and the parameter $q$. This paper will show that given a random variable with a probability density of $f(x)$, integer values of $q$ quantify the number of random variables required to be equal in forming a subsample that is distributed as the escort density. And that approximates of these subsamples, referred to as *Independent Approximates* (IAs), enable closed-form estimation of heavy-tailed distributions. While not completed here, it is conjectured that the properties described would also extend to non-integer values of $q$.

The process of selecting IA pairs and triplets is illustrated in Fig. 1 for a Cauchy distribution. Given independent, identically distributed samples from a one-dimensional distribution, the IAs are formed by first forming an $n^{th}$-order distribution through the random selection of pairs, triplets, or higher groups. Groups that are approximately equal are selected within a tolerance bound of the diagonals. Within each approximately equal grouping, the median is retained as a subsample. Evident from the figure is the fact that the subsamples along the diagonal have a significantly faster rate of decay in the tails then original samples along each axis. In Section 3.2 a proof is provided that marginal probability density function (pdf) along the diagonal is the normalized $n^{th}$-*power-density* of the original pdf $f_X(x)$,

$$f_X^{(n)}(x) \stackrel{\text{def}}{=} {f_X^n(x)}\Big/{\int_{x \in X} f_X^n(x)}.$$

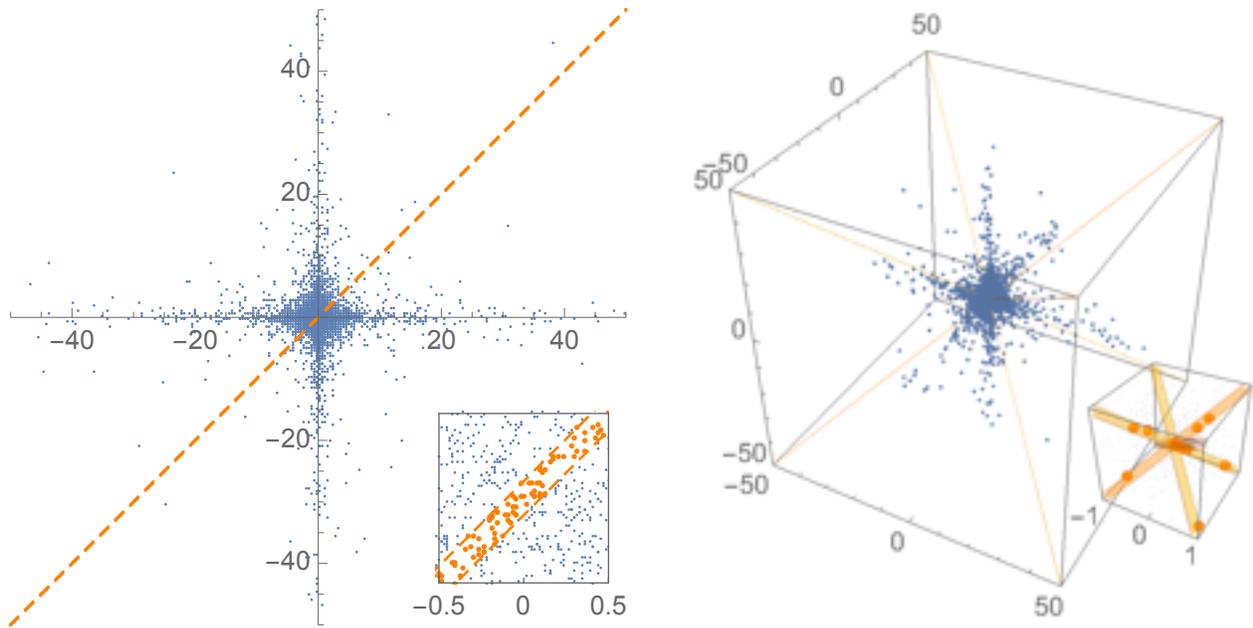

**Fig. 1** Selection of Independent Approximates from a standard Cauchy distribution. a) Pairs are selected along the equal diagonal within a range of 0.1 and are used to estimate the first-moment. b) Triplets are selected along all the diagonals since these samples will be used to estimate the second-moment which is invariant to the sign. The range from -50 to 50 shows how the distribution along the diagonals has significantly fewer outliers due to the faster tail decay. The insets show the selected samples near the origin.

The paper will show that pairs of IAs ensures that the first moment of heavy-tailed distributions is defined, triplets ensure that the second moment is defined, and that higher-order IAs ensure the definition of the moment one less than the partition size of the IAs. And importantly, there is a functional mapping between the moments of the IAs and the parameters of the original distribution. The performance of estimators is examined both theoretically and empirically. The estimator is shown to approximate the maximum likelihood estimator for small sample sets and converge to precise estimation for large sample sets. Trade-offs in the performance include 1) while the moments are defined, the domain of finite variance of the estimators is limited, 2) a tolerance for accepting approximates must balance filtering unequal samples with choosing an adequate subsample set for statistical analysis, and 3) use of iterative perturbations for selecting the grouping of independent samples can increases the subsample size but distorts the subsequent distribution and decreases the speed of the algorithm.

In the next section, the relationship between nonlinearity, $q$-statistics, and more traditional characterizations of traditional characterizations of heavy-tailed distributions and their estimators are reviewed. In Section 3, a formal definition of Independent Approximates is established along with their relationship to the powers of probability density functions. In Section 4, theoretical derivations of the power distributions for the generalized Pareto and Student's $t$ are completed. Section 5 documents the computational estimation of the Student's t-distribution using the IA algorithm. Section 6 explores both small and large sample applications. Section 7 gives a concluding discussion regarding the significance of the methodology and future directions for the research.

## 2. Review of heavy-tailed estimation methods

### 2.1. Complex systems and heavy-tailed distributions

Mathematicians, engineers, and scientists have for over a hundred years sought to characterize the statistics of complex systems using a variety of perspectives regarding heavy-tailed distributions. Despite recent alternatives, the frameworks defined by Vilfredo Pareto (Pareto 1896)and William Gosset (Gossett 1904) have important and under-appreciated advantages in the modeling of complex systems. Paul Lévy's (Lévy 1948) characterization of stable distributions, while providing an important mathematical framework, requires an assumption that the underlying phenomena starts with heavy-tailed behavior. In contrast, the work of Alfréd Rényi (Rényi and Makkai-Bencsáth 1984) and Constantino Tsallis (Tsallis 2009) showed that nonlinear dynamics lead to convergence toward heavy-tailed distributions. Unfortunately, the framework now referred to as nonextensive statistical mechanics, does not specify the physical property represented by the parameter $q$. This paper clarifies that given the origination of $q$-entropy in the examination of systems characterized by $f^q(x) / \int f^q(x)$, that this power density represents a subsample in which random variables share a common state. Further, the paper will show that $q$ can be decomposed into more fundamental properties, including the shape parameter. And that the shape parameter can complement information measures of complexity (López-Ruiz et al. 1995; Standish 2008) by specifying the portion due to nonlinear dynamics and heavy-tailed statistics. which could be a candidate for quantifying complexity.

To begin a mathematical review of the Pareto and Student's t distributions will provide a grounding in the essential parameters of location, scale, and shape of a heavy-tailed distribution. Estimation of the skew using IAs is deferred for future research. Both the Pareto and Student's t distributions are characterized by a shape parameter, the tail index ($\alpha$) and the degree of freedom ($\lambda$), respectively. The inverse of these parameters, also called the shape parameter ($\kappa$) turns out to be a natural candidate for quantifying complexity, since at zero the tail decay is exponential (i.e. no nonlinearity resulting in complexity) and approaching infinity the underlying correlations are extreme. In my own prior research, I have referred to this shape parameter as the *Nonlinear Statistical Coupling* (Nelson and Umarov 2010; Nelson et al. 2017), since it can be related to sources of nonlinearity, such as multiplicative noise or fluctuations in the variance. For the sake of grounding the discussion in traditional statistical methods, I'll refer to this as the shape parameter.

The Type II Pareto distribution has a survival function (sf) defined by

$$1 - F\left(x\right) = \begin{cases} \left(1 + \kappa \dfrac{x-\mu}{\sigma}\right)^{-1/\kappa} & \kappa \neq 0, x \geq \mu \\ \\ e^{-\frac{x-\mu}{\sigma}} & \kappa = 0, x \geq \mu, \end{cases} \qquad (2.1)$$

where $F$ is the cumulative distribution function (CDF), $\mu \in (-\infty, \infty)$ is the location, $\sigma > 0$ is the scale, and $0 \leq \kappa \leq \infty$ is the shape. A compact-support domain $-1 \leq \kappa < 0$ with faster than exponential decay can also be defined but is not the focus of this paper. The exponential case is $\kappa = 0$. For the heavy-tailed domain, the generalized Pareto pdf is

$$f\left(x\right) = F'\left(x\right) = \frac{1}{\sigma}\left(1 + \kappa\frac{x-\mu}{\sigma}\right)^{-\frac{1+\kappa}{\kappa}}, \kappa > 0. \tag{2.2}$$

In the limit where $x \to \mu\kappa$, the Type II converges to the purely power-law Type I Pareto

$$1 - F(x) = \left(\frac{x}{\mu}\right)^{-\frac{1}{\kappa}}, \ x \geq \mu. \tag{2.3}$$

In the limit where $\mu \to 0$, the expression scale-free has an actual mathematical description in which the scale has gone to zero. For the remainder of the paper, the generalized (Type II) Pareto distribution refers to the Pareto distribution.

Likewise, with the Student's $t$ an adjustment is made in the traditional specification of the distribution. In earlier research, the author showed that the shape parameter of heavy-tailed distributions is equal to a source of nonlinearity (Nelson and Umarov 2010; Nelson et al. 2017), e.g. the magnitude of fluctuations in superstatistics and the multiplicative noise in certain stochastic processes (Sornette 1998). For this reason, the shape is also referred to as the nonlinear statistical coupling or simply the coupling; and a family of distributions referred to as the Coupled Exponential Family (Nelson 2015) has been defined. In this paper, the distributions (Pareto and Student's $t$) will be referred to by their traditional names, but the insights regarding the importance of the shape parameter will be emphasized. Thus, the Student's $t$ will be defined using the shape parameter which is the reciprocal of the degree of freedom $\left(v = \frac{1}{\kappa}\right)$. The sf is

$$1 - F\left(x\right) = -\frac{1}{2} - \frac{x\sqrt{\kappa}}{\sigma B\left(\frac{1}{2\kappa},\frac{1}{2}\right)}{}_2F_1\left(\frac{1}{2},\frac{1+\kappa}{2\kappa},\frac{3}{2};-\kappa\frac{\left(x-\mu\right)^2}{\sigma^2}\right), \ \sigma \geq 0, \ \kappa > 0, \tag{2.4}$$

where ${}_2F_1$ is the hypergeometric function and the pdf is

$$f\left(x;\mu,\sigma,\kappa\right) = \frac{\sqrt{\kappa}}{\sigma B\left(\frac{1}{2\kappa},\frac{1}{2}\right)}\left(1 + \kappa\frac{\left(x-\mu\right)^2}{\sigma^2}\right)^{-\frac{1}{2}\left(\frac{1}{\kappa}+1\right)}, \ \sigma \geq 0, \ \kappa > 0. \tag{2.5}$$

Not all heavy-tailed distributions have an analytical expression for the pdf. An important example is the Lévy-Stable distribution, which is instead defined by its characteristic function, a generalized Gaussian

$$\phi_X\left(x\right) = \exp\left(it\mu - \left|\sigma t\right|^\alpha\right), \ 0 < \alpha \leq 2 \tag{2.6}$$

where $\mu$ is the location, $\sigma$ is the scale, $\alpha$ is the stability parameter that affects the shape, and the skew parameter has been set to zero. The Fourier transform of (2.6) defines the pdf

$$f_X\left(x\right) = \frac{1}{2\pi}\int_{-\infty}^{\infty}\phi\left(t\right)e^{-ixt}\,dt. \tag{2.7}$$

The asymptotic tail of the Levy-stable distribution is $\left|x\right|^{-(\alpha+1)}$, thus the shape is $\kappa = \frac{1}{\alpha}$ and has a domain of $\kappa \geq \frac{1}{2}$, i.e. the domain of infinite variance. While it is anticipated that IAs could be used for estimation of the Lévy-Stable distributions, the analysis requires consideration of power-convolutions and is deferred for future research.

## 2.2. Estimation of heavy-tailed distributions

Much of the scientific literature related to complex systems and their underlying heavy-tailed distributions has focused on non-exponential tail decay while treating the main body of the distribution as a separate function. As such, many of the existing estimation methods focus exclusively on estimates of the power-law or Pareto type I tail. Fedotenkov (Fedotenkov 2018) provides an extensive review of estimation methods for Pareto or power-law tail decay. Many of the estimators reviewed by Fedotenkov are refinements of the maximum likelihood Hill estimator (Hill 1975) which compares the logarithm of the highest order statistics $X_{(k)}....X_{(n)}$

$$\hat{\kappa}_n^H\left(k\right) = \frac{1}{k}\sum_{i=0}^{k-1}\log\left(\frac{X_{(n-i)}}{X_{(n-k)}}\right). \tag{2.8}$$

A key component of the Hill and related estimators is the need to choose $k$ such that it is a) sufficiently greater than the location plus the scale $\mu+\sigma$ so that remaining order statistics are approximately power-law, i.e. linear on a log-log scale, and b) sufficiently smaller than n to provide an adequate number of samples. The Hill estimate is based on measuring the average distance on the log scale between the $n-i$ order statistic and the $n-k$ order statistic provides an estimate of the tail-index. Advanced variations (Stoev et al. 2011) expand upon this relationship to improve the regression analysis of the log-log relationship of the tail decay. A recent example that iteratively optimizes both the cut-off and the estimate is reported by Hanel, et al. (Hanel et al. 2017).

The nonextensive statistical mechanics community has focused on a family of distributions ($q$-statistics) that include the generalized Pareto and Student's $t$ distributions. Through a generalization of the logarithm and exponential functions that accounts for the non-exponential tail-decay estimator, methods have been developed that maximize the generalized log-likelihood (Shalizi 2007; Ferrari and Yang 2010; Nielsen 2013; Qin and Priebe 2013; Gayen and Kumar 2018). The generalized likelihood method weights the gradient of the log-likelihood by the power of the pdf (Ferrari and Yang 2010)

$$\sum_{i=1}^{N} f_X^{\frac{-\kappa}{1+\kappa}} \nabla_\theta \log f_X\left(x_i, \theta\right) = 0, \tag{2.9}$$

where $\theta$ is the location, scale, and possibly other non-shape parameters of the distribution. The generalized log-likelihood methods depend on a separate estimate of the tail such as the Hill-like estimators.

Further, (Tsallis et al. 2009) showed that generalized moments based on $q$-statistics correspond to the parameters of the generalized Pareto and Student's t distributions. The exponent of (2.5), in which 2 is generalized to $\alpha$, defines the parameter $q$ via the relationship

$$\frac{1}{q-1} = \frac{1+\kappa}{\alpha\kappa} \text{ or } q = 1 + \frac{\alpha\kappa}{1+\kappa}. \tag{2.10}$$

The $q$ is dependent on the shape parameter $\kappa$ and the power of the variable, $\alpha$, which is one for the Pareto distributions and two for the Student's $t$-distribution. Rather than a scale, $q$-statistics typically uses $\beta$ which is defined by the multiplicative term of the variable in (2.5),

$$\beta(1-q) = \frac{\kappa}{\sigma^\alpha} \text{ or } \beta = \frac{\kappa}{(1-q)\sigma^\alpha}. \tag{2.11}$$

These generalized moments are defined as

$$E[X^m; \alpha, \kappa] = \frac{\int_{x \in X} x^m f_X^{1+\frac{m\alpha\kappa}{1+\kappa}}(x)dx}{\int_{x \in X} f_X^{1+\frac{m\alpha\kappa}{1+\kappa}}(x)dx}. \qquad (2.12)$$

Thus, $q$ defines the power of the first generalized moment ($m = 1$) and therefore defines the grouping of a fractional number of random variables into the same state to form the generalized moment. And the shape parameter whose heavy-tail domain is 0 to $\infty$ is a more natural candidate for quantifying the portion of complexity due to nonlinearity.

Nevertheless, the focus of this paper will be on using integer powers of the distribution, thus eliminating the need to match the generalized moments with the properties of the distribution. In prior research, the author with Umarov and Kon (Nelson et al. 2019) showed that given the location of a Student's $t$-distribution a functional relationship exists between the scale and shape parameters using the log-average or geometric mean of the samples

$$\hat{\sigma} = 2\sqrt{\kappa} \exp\left(\frac{1}{2}H_{\left(-1+\frac{1}{2\kappa}\right)}\right)\prod_{i=1}^{N}|x_i - \mu|^{1/N}, \qquad (2.13)$$

where $H_n$ is the $n^{th}$ harmonic number. Given the new estimators for the location and scale defined in the next section, this geometric mean estimator will be used to estimate the shape of the Student's $t$. In the context of IAs, this estimator can be regarded as the $0^{th}$ moment of the 1 power-distribution (i.e. no subsampling is necessary) and will be used to estimate the shape given estimates of the location and scale.

## 3. Independent Approximates

### 3.1. Significance of Power-Distributions

Heavy-tailed distributions have a limited number of moments, and potentially no moments, which are both defined and finite. The moments $\mu_m$ are defined by the derivative at $t = 0$ of the characteristic function $\phi$ and equivalently the moment-generating function $M$

$$\mu_m = E\left[X^m\right] = M_X^{(m)}\left(0\right) = i^m \phi_X^{(m)}\left(0\right) = i^{-m}\frac{d^m E\left[e^{itX}\right]}{dt^m}\Bigg|_{t=0} \qquad (3.1)$$

Given a heavy-tailed distribution with an asymptotic power between $1 < 1 + \nu < \infty$ or equivalently using the shape parameter $1 < 1 + \frac{1}{\kappa} < \infty$, raising the distribution to a power greater than 1 increases the rate of decay of the distribution thereby increasing the number of moments that are defined and finite. So, for instance, the standard Cauchy density $\pi^{-1}\left(1+x^2\right)^{-1}$ has a characteristic function $e^{-|t|}$ whose derivatives at $t = 0$ are discontinuous; thus, the moments are not defined. Raising the Cauchy density to the $n^{th}$ power $\pi^{-n}\left(1+x^2\right)^{-n}$ results in a Student's t density upon normalization. The shape of this power-density is $\kappa = \frac{1}{n-1}$ which is less than 1 for $n > 2$, and would then have a defined first moment, and is less than 0.5 for $n > 3$, and would then have a finite second moment.

### 3.2. Selecting Independent Approximate Subsamples

By itself raising a density to a power does not lead to a closed-form estimator of the distribution because the empirical distribution would first have to be estimated prior to applying the power. This circularity requires iterative methods for estimation. Nevertheless, a closed-form estimation can be achieved by noticing that given a joint distribution with $n$ independent dimensions, the marginal distribution along the diagonal with equal values has the required $n^{th}$-power-density.

**Lemma 1:** *Given an n-dimensional random variable $X$ with identical and independent marginal densities equal to $f_{X_i}(x_i)$, the marginal density along the diagonal of n equal values, has a density equal to $f_X^{(n)}(x) = f^n(x) \bigg/ \int_{x \in X} f^n(x)$.*

Proof: The marginal distribution along the equally-valued diagonal is a normalization of the joint distribution with all $x_i = x$

$$f_{\mathbf{X}}(\mathbf{x})\Big|_{x_i = x} = \prod_{i=1}^{n} f_X(x_i)\Big|_{x_i = x} = f_X^n(x). \tag{3.2}$$

Renormalizing the result to form the marginal density completes the proof. □

**Definition 1** (Independent-Equals) *Random samples drawn from the marginal distribution of an n-dimensional random variable $\mathbf{X}$ with independent dimensions are defined to be Independent-Equals and are symbolically represented as $X^{(n)} \sim f^{(n)}(x)$.*

**Definition 2** (Independent-Approximates) *Random samples drawn from a neighborhood of tolerance $\varepsilon$ of the independent-equals marginal distribution are defined as Independent-Approximates (IAs). The IAs samples are also represented as $X^{(n)} \sim f^{(n)}(x)$ with the distinction between approximate and equal being described in the context of either theoretical computations assuming equality or experimental computations implementing approximate methods. The neighborhood for selection of IAs is defined as $\left| x_i^{(n)} - x_j^{(n)} \right| \leq \varepsilon$ where $x_i$ are samples from the n independent dimensions or equivalently partitions with n independent samples.*

Thus, estimators for the $n^{th}$ power of a density can be formed by partitioning $N$ independent, identically distributed samples into subsets of length $n$ and then selecting those subsets which are approximately equal. The median of those subsets forms a set of *Independent Approximates* (IA) whose size will be labeled $N^{(n)}$. Furthermore, for even moments the required equality is for the absolute value of x, expanding the diagonals which can be utilized for the estimator.

**Corollary 1:** *Given n independent, identically distributed random variables with density f, the marginal distribution of equal absolute value has a density equal to* $f_X^{(n)}\left(|x|\right) = f^n\left(|x|\right) \Big/ \int_{x \in X} f^n\left(|x|\right)$.

The proof has the same structure of Theorem 1 with $x$ replaced by $|x|$.

**Fig. 1** illustrates the selection of pairs and triplets of *Independent Approximates* (IA) for a standard Cauchy random variable. A tolerance of $\pm 0.1$ has a reasonable balance between filtering outliers and selecting approximates to the diagonal. As will be described in detail in the next section the pairs are selected along the diagonal of equal value and are used to measure the first moment. The triplets are selected along all the diagonals since their use for estimating the second moment is insensitive to the sign of numbers. The selected set of IAs becomes one sample, the mean of each pair and the median of each triplet. In the next section, power moments are derived for the generalized Pareto and Student's *t* distributions.

## 4. Power-density moments

The power-density moments or power-moments provide a mapping between estimates that can be measured given the reduction in the shape of the distribution and the original distribution.

**Definition 3: (Power-moment)** *The $m^{th}$ moment of the $n^{th}$ power of density $f_X\left(x\right)$ is defined as*

$$\mu_m^{(n)} \underset{\text{def}}{=} \frac{\int_{x \in X} x^m f_X^n(x)dx}{\int_{x \in X} f_X^n(x)dx} = E\left[\left(X^{(n)}\right)^m\right] \tag{4.1}$$

### 4.1. The generalized Pareto distribution

**Lemma 2:** *Given a one-sided (two-sided) generalized Pareto distribution with density*

$$f_X\left(x\right) = \frac{1}{\sigma}\left(1 + \kappa\frac{x-\mu}{\sigma}\right)^{-\left(\frac{1}{\kappa}+1\right)}; x \geq \mu$$

$$f_X\left(x\right) = \frac{1}{\sigma}\left(1 + \kappa\left|\frac{x-\mu}{\sigma}\right|\right)^{-\left(\frac{1}{\kappa}+1\right)}; -\infty \leq x \leq \infty \tag{4.2}$$

*the nth moment exists and is finite for all shapes $\left(\kappa \geq 0\right)$ if the distribution is raised to the $(n+1)-$ power and renormalized. The functional relationship between moments for the $n^{th}$ power of the generalized Pareto distribution and the location, scale, and shape parameters are specified in **Table 1**(**Table** 2).*

**Table 1:** The nᵗʰ moments for the (n+1)-power-density of the one-sided generalized Pareto distribution.

| Moment, Centered | One-sided Pareto Type II | |
| --- | --- | --- |
| | Non-centered | Centered |
| $\mu_1^{(2)}, x-\mu$ | $\mu+\dfrac{\sigma}{2}$ | $\dfrac{\sigma}{2}$ |
| $\mu_2^{(3)}, x-\mu$ | $\mu^2+\dfrac{2\mu\sigma}{3+\kappa}+\dfrac{2}{3}\dfrac{\sigma^2}{3+\kappa}$ | $\dfrac{2\sigma^2}{3(3+\kappa)}$ |
| $\mu_3^{(4)}, x-\mu$ | $\mu^3+\dfrac{3\mu^2\sigma}{(4+2\kappa)}+\dfrac{12\mu\sigma^2+3\sigma^3}{2(4+\kappa)(4+2\kappa)}$ | $\dfrac{3\sigma^3}{2(4+\kappa)(4+2\kappa)}$ |
| $\mu_4^{(5)}, x-\mu$ | $\mu^4+\dfrac{4\mu^3\sigma}{5+3\kappa}+\dfrac{12\mu^2\sigma^2}{(5+2\kappa)(5+3\kappa)}+\dfrac{120\mu\sigma^3+24\sigma^4}{5(5+\kappa)(5+2\kappa)(5+3\kappa)}$ | $\dfrac{24\sigma^4}{(5+\kappa)(5+2\kappa)(5+3\kappa)}$ |
| $\mu_n^{(n+1)}, x-\mu$ | $\displaystyle\sum_{i=0}^{n}\dfrac{n!\mu^{n-i}\sigma^i}{(n-i)!}\dfrac{\frac{1+n+(n-i-1)\kappa}{\kappa}!}{(\kappa^i)\frac{1+n+(n-1)\kappa}{\kappa}!}$ | $\kappa^{-n}\dfrac{n!\frac{1+n-\kappa}{\kappa}!}{\frac{1+n+n\kappa-\kappa}{\kappa}!}\sigma^n$ |

**Table 2:** The nᵗʰ moments for the (n+1)-power-density of the two-sided generalized Pareto distribution.

| Moment, Centered | Two-sided Pareto Type II | |
| --- | --- | --- |
| | Non-centered | Centered |
| $\mu_1^{(2)}, x-\mu$ | $\mu$ | $0$ |
| $\mu_2^{(3)}, x-\mu$ | $\mu^2+\dfrac{4}{3}\dfrac{\sigma^2}{3+\kappa}$ | $\dfrac{4\sigma^2}{3(3+\kappa)}$ |
| $\mu_3^{(4)}, x-\mu$ | $\mu^3+\dfrac{12\mu\sigma^2+3\sigma^3}{(4+\kappa)(4+2\kappa)}$ | $\dfrac{3\sigma^3}{(4+\kappa)(4+2\kappa)}$ |
| $\mu_4^{(5)}, x-\mu$ | $\mu^4+\dfrac{24\mu^2\sigma^2}{(5+2\kappa)(5+3\kappa)}+\dfrac{48\sigma^4}{5(5+2\kappa)(5+3\kappa)}$ | $\dfrac{48\sigma^4}{(5+\kappa)(5+2\kappa)(5+3\kappa)}$ |
| $\mu_n^{(n+1)}, x-\mu$ | $\displaystyle\sum_{i=0}^{n}\left(1+(-1)^n\right)\dfrac{n!\mu^{n-i}\sigma^i}{(n-i)!}\dfrac{\frac{1+n+(n-i-1)\kappa}{\kappa}!}{(\kappa^i)\frac{1+n+(n-1)\kappa}{\kappa}!}$ | $\left(1+(-1)^n\right)\dfrac{\kappa^{-n}n!\frac{1+n-\kappa}{\kappa}!}{\frac{1+n+n\kappa-\kappa}{\kappa}!}\sigma^n$ |

*Proof:* The normalized $(n+1)$-power-density of a single-sided generalized Pareto density (Type II) is also a Pareto density with a modified scale and shape

$$f_X^{(n+1)}(x) = \frac{\dfrac{1}{\sigma^{n+1}}\left(1+\kappa\dfrac{x-\mu}{\sigma}\right)^{-\left(\frac{1}{\kappa}+1\right)(n+1)}}{\displaystyle\int_\mu^\infty \dfrac{1}{\sigma^{n+1}}\left(1+\kappa\dfrac{x-\mu}{\sigma}\right)^{-\left(\frac{1}{\kappa}+1\right)(n+1)}dx} \ . \tag{4.3}$$

Equating the new exponent with the structure of the generalized Pareto distribution determines the modified shape

$$\left(\frac{1}{\kappa'}+1\right)=\left(\frac{1}{\kappa}+1\right)(n+1)$$

$$\kappa'=\frac{\kappa}{1+n+n\kappa}. \tag{4.4}$$

Equating the factors multiplying the variable determines the modified scale

$$\frac{\kappa'}{\sigma'}=\frac{\kappa}{\sigma}$$

$$\sigma'=\frac{\sigma}{1+n+n\kappa}. \tag{4.5}$$

The $n^{th}$ moment of the Pareto distribution is restricted to the domain $\kappa < 1/n$, thus for the $n+1$ power-distribution is the region of finite existence for the moments is

$$\kappa'=\frac{\kappa}{1+n+n\kappa}<\frac{1}{n}$$

$$0<\frac{1}{n}+1. \tag{4.6}$$

Thus, the moment for the n+1 power-density of the Pareto distribution exists and is finite for all values of the shape parameter. Solutions for the moment integrals

$$\mu_n^{(n+1)}=\frac{\displaystyle\int_\mu^\infty x^n\frac{1}{\sigma}\left(1+\kappa\frac{x-\mu}{\sigma}\right)^{-\left(\frac{1}{\kappa}+1\right)(n+1)}dx}{\displaystyle\int_\mu^\infty \frac{1}{\sigma}\left(1+\kappa\frac{x-\mu}{\sigma}\right)^{-\left(\frac{1}{\kappa}+1\right)(n+1)}dx} \tag{4.7}$$

are specified in **Table 1**. The $n+1$ power-distribution for the two-sided generalized Pareto has the same modified shape and thus the $n^{th}$ also exists for all shape values. The solutions for the two-sided Pareto are specified in **Table 2**. □

**Corollary 2.1** *Given the location of a one-sided generalized Pareto distribution, the scale is determined independently of the shape by the first moment with the pdf raised to the second power and normalized.*

$$\sigma = 2\mu_1^{(2)} = 2E^{(2)}\big[X - \mu\big] = 2\frac{\int_0^\infty x\left(1 + \kappa\frac{x}{\sigma}\right)^{-\left(\frac{1}{\kappa}+1\right)2} dx}{\int_0^\infty \left(1 + \kappa\frac{x}{\sigma}\right)^{-\left(\frac{1}{\kappa}+1\right)2} dx} \tag{4.8}$$

**Corollary 2.2** *Given the location and scale of a one-sided generalized Pareto distribution, the shape is determined by second moment with pdf raised to the third power and normalized.*

$$\kappa = \frac{4}{3\mu_2^{(3)}} - 3 = \frac{4}{3E^{(3)}\Big[\left(X-\mu\right)^2\Big]} - 3 = \frac{4\int_0^\infty \left(1 + \kappa\frac{x}{\sigma}\right)^{-\left(\frac{1}{\kappa}+1\right)3} dx}{3\int_0^\infty x^2 \left(1 + \kappa\frac{x}{\sigma}\right)^{-\left(\frac{1}{\kappa}+1\right)3} dx} - 3 \tag{4.9}$$

**Corollary 2.3** *The location of a two-sided generalized Pareto distribution is determined by the first moment with the pdf raised the second power and normalized.*

$$\mu = \mu_1^{(2)} = E^{(2)}\big[X\big] = \frac{\int_{-\infty}^\infty x\left(1 + \kappa\left|\frac{x-\mu}{\sigma}\right|\right)^{-\left(\frac{1}{\kappa}+1\right)2} dx}{\int_{-\infty}^\infty \left(1 + \kappa\left|\frac{x-\mu}{\sigma}\right|\right)^{-\left(\frac{1}{\kappa}+1\right)2} dx} \tag{4.10}$$

**Corollary 2.4** *Given the location of a two-sided generalized Pareto distribution the scale is determined independently of the shape by applying Corollary 1.1 to the $x > \mu$ side of the distribution* (4.8). *Likewise, the shape is determined by applying Corollary 1.2 to the $x > \mu$ side* (4.9).

### 4.2. The Student's t distribution

**Lemma 3:** *Given a non-centered and scaled Student's t distribution* (2.5) *the generalized nth moment exists for all shapes $\left(\kappa \geq 0\right)$ for the $\left(n+1\right)$-power-distribution. The generalized moments for the Student's t distribution are specified in Table 3.*

*Proof:* The normalized $\left(n+1\right)$-power-density of a Student's *t*-density is also a Student's *t* with a modified scale and shape, where the shape is the inverse of the degree of freedom. The power-density, neglecting the original normalization which cancels out, is

$$f_X^{(n+1)}\left(x\right) = \frac{\left(1 + \kappa\left|\frac{x-\mu}{\sigma}\right|^2\right)^{-\frac{1}{2}\left(\frac{1}{\kappa}+1\right)\left(n+1\right)}}{\int_{-\infty}^\infty \left(1 + \kappa\left|\frac{x-\mu}{\sigma}\right|^2\right)^{-\frac{1}{2}\left(\frac{1}{\kappa}+1\right)\left(n+1\right)} dx} \quad . \tag{4.11}$$

**Table 3:** The n$^{th}$ moments for the (n+1)-power-density the Student's $t$-distribution.

| Moment, Centered | Student's t, n+1 | |
|:---:|:---:|:---:|
| | Non-centered | Centered |
| $\mu_1^{(2)}$ | $\mu$ | - |
| $\mu_2^{(3)}, x-\mu$ | $\mu^2 + \dfrac{\sigma^2}{3}$ | $\dfrac{\sigma^2}{3}$ |
| $\mu_3^{(4)}, x-\mu$ | $\mu^3 + \dfrac{3\mu\sigma^2}{4+\kappa}$ | $0$ |
| $\mu_4^{(5)}, x-\mu$ | $\mu^4 + \dfrac{6\mu^2\sigma^2}{5+2\kappa} + \dfrac{3\sigma^4}{5(5+2\kappa)}$ | $\dfrac{3\sigma^4}{25+10\kappa}$ |
| $\mu_n^{(n+1)}, x-\mu$ | Simplification Not Available | $\left(1+(-1)^n\right)\left(\dfrac{\sigma}{\sqrt{\kappa}}\right)^n \dfrac{\frac{n-1}{2}! \frac{n+1-2\kappa}{2\kappa}!}{2\sqrt{\pi}\,\frac{n+1+(n-2)\kappa}{2\kappa}!}$ |

The modified shape parameter is the same as the Pareto power-distribution since the $\frac{1}{2}$ factor cancels out

$$\left(\frac{1}{\kappa'}+1\right) = \left(\frac{1}{\kappa}+1\right)(n+1)$$

(4.12)

$$\kappa' = \frac{\kappa}{1+n+n\kappa}.$$

The modified scale is determined by the terms multiplying the variable

$$\frac{\kappa'}{\sigma'^2} = \frac{\kappa}{\sigma^2}$$

(4.13)

$$\sigma' = \frac{\sigma}{\sqrt{1+n+n\kappa}}.$$

Like the generalized Pareto distribution, existent, finite $n^{th}$ moments of the Student's $t$-distribution are restricted to the domain $\kappa < \frac{1}{n}$, thus from Lemma 2 the $n+1$ power-density of the Student's $t$ has a finite $n^{th}$ finite moment for all shapes. Solutions for the moment integrals

$$\mu_n^{(n+1)} = \frac{\displaystyle\int_{-\infty}^{\infty} x^n \left(1+\kappa\left(\frac{x-\mu}{\sigma}\right)^2\right)^{-\frac{1}{2}\left(\frac{1}{\kappa}+1\right)(n+1)} dx}{\displaystyle\int_{-\infty}^{\infty} \left(1+\kappa\left(\frac{x-\mu}{\sigma}\right)^2\right)^{-\frac{1}{2}\left(\frac{1}{\kappa}+1\right)(n+1)} dx}$$

(4.14)

are specified in **Table 3.** □

General $m^{th}$-moment and $n$-power-density $\mu_m^{(n)}$ serve a variety of purposes. In the next section, the variance of estimators will be shown to depend on twice the moment of the estimator $\mu_{2m}^{(n)}$. Here it is noted that further increments of the power-density, such as $\mu_n^{(n+2)}$ shown in **Table 4**, provide additional relationships between the power-moments and distribution parameters. The general power-moments for the Student's $t$-distribution was computed using *Mathematica*

$$\mu_m^{(n)} = \int_{-\infty}^{\infty} x^m f^n(x)\,dx \,/\, \int_{-\infty}^{\infty} f^n(x)\,dx$$

$$= \frac{\left(1+\left(-1\right)^m\right)\kappa^{-m/2}\sigma^m\left(\dfrac{m-1}{2}\right)!\left(\dfrac{n+\left(n-m-3\right)\kappa}{2\kappa}\right)!}{2\sqrt{\pi}\left(\dfrac{n+\left(n-3\right)\kappa}{2\kappa}\right)!}\ \text{ for }\kappa<\frac{n}{1+m-n}. \tag{4.15}$$

From this general solution, the infinite domain when $n = m+1$ is evident.

**Table 4:** The $n^{th}$ moments for the (n+2)-power-density of the Student's $t$-distribution.

| Moment, Centered | Student's t, n+2 | |
|---|---|---|
| | Non-centered | Centered |
| $\mu_1^{(3)}$ | $\mu$ | - |
| $\mu_2^{(4)},\ x-\mu$ | $\mu^2 + \dfrac{\sigma^2}{4+\kappa}$ | $\dfrac{\sigma^2}{4+\kappa}$ |
| $\mu_3^{(5)},\ x-\mu$ | $\mu^3 + \dfrac{3\mu\sigma^2}{5+2\kappa}$ | $0$ |
| $\mu_4^{(6)},\ x-\mu$ | $\mu^4 + \dfrac{2\mu^2\sigma^2}{2+\kappa} + \dfrac{\sigma^4}{\left(2+\kappa\right)\left(6+\kappa\right)}$ | $\dfrac{\sigma^4}{\left(6+\kappa\right)\left(2+\kappa\right)}$ |
| $\mu_n^{(n+2)},\ x-\mu$ | Simplification Not Available | $\left(1+\left(-1\right)^n\right)\left(\dfrac{\sigma}{\sqrt{\kappa}}\right)^n \dfrac{\frac{n-1}{2}!\frac{n+2-\kappa}{2\kappa}!}{2\sqrt{\pi}\frac{n+2+\left(n-1\right)\kappa}{2\kappa}!}$ |

**5. Performance of the IA algorithm for estimation of the Student's *t*-distribution**

The Student's *t*-distribution is used to demonstrate and evaluate the performance of the *Independent Approximates* algorithm. Because the subsampling decimates the samples by a factor of about a thousand, a baseline of 10,000 samples with 10 iterations of partitioning and selecting approximates is used. The generalized Box-Müller method (Nelson and Thistleton 2006 May 23) is used to generate random samples. So that the bias, precision, and accuracy of the estimators can be measured each estimate is computed twenty times. First, a more detailed description of the algorithm is provided.

**5.1. Description of the IA Algorithm**

*Mathematica* software implementing the IA algorithm is available at the Photrek Github repository (Nelson 2020a) and calls functions from the Nonlinear-Statistical-Coupling repository (Nelson 2020b). The program flow of the algorithm to compute the parameters of a heavy-tailed distribution consists of:

1) center the samples using the median and scale the samples using the 75% quantile minus the median. This ensures the tolerance selection will be consistent for a variety of distributions.
2) permute the samples and select IAs several times (e.g. 10) by
   a. If estimating location
      i. partition samples into pairs; (optional: with an overlapping offset fewer IA pairs will be missed but pairs sharing a sample will be correlated)
      ii. select pairs which are equal within a tolerance (e.g. 0.1), **Fig. 1a**
   b. If estimating scale
      i. partition samples into triplets; (optional: with an overlapping offset fewer IA triplets will be missed but triplets sharing a sample will be correlated)
      ii. since the $2^{nd}$ moment is insensitive to the sign of samples, approximates can be selected along three diagonals, however, each diagonal must be searched separately since taking the absolute value first creates a bias of values of not near the origin, **Fig. 1b**
   c. If estimating a higher moment, partition into groups of *n* and select either along the equal diagonal for an odd moment or along all the diagonals if an even moment
3) use the IA subsamples to compute the required moment of the power-distribution
4) compute the desired parameter of the original distribution. For the generalized Pareto or Student's *t,* the relationships are defined in **Table 1** thru **Table 4**.

The selection of IAs is sensitive to the tolerance criteria for choosing partitions with approximately equal samples. Based on experiments with 0.01, 0.1, and 1 as tolerance parameters, a tolerance of 0.1 provides a good balance between selecting a sufficient number of samples while ensuring that the desired filtering for a distribution approximating $f^n$ is achieved. Additional research is planned to quantify this trade-off. The 10 permutations to select IAs takes approximately 0.2 seconds for the pairs, and approximately 2 seconds for the triplets running in *Mathematica* on a MacBook Pro with a 2.3 GHz Intel Core i5 and 16 GB of memory. Using the *Mathematica* CloudEvaluation, which computes the computation of a 64-bit Linux x86 server, takes approximately 0.35 seconds for both selections. Given the equal speeds, the latency of the cloud is presumably a significant factor.

There are several options for estimating the shape of the Student's *t* given estimates of the location and scale. For this study, the geometric mean estimator (2.13) defined by Nelson et. al (Nelson et al. 2019) will be used. Although this is not strictly a closed-form estimate, the root finder algorithms required are well established. Both use of the 4-power-density to measure $\mu_2^{(4)}$ and the Hill-type estimators are alternatives that could provide a closed-form estimator. Comparative performance of the speed and accuracy of these and other shape estimators will be explored in future research.

### 5.2. Expected Bias and Precision

The accuracy (root mean square error) of an estimate is a function of the bias (expected error of the estimate) and the precision (standard deviation of the estimates). While the $(n+1)-$power-density is sufficient to define the $n^{th}$ moment and thus assures the bias will be finite, this does not assure that the precision is finite. Since the variance of the $n^{th}$ moment estimator is proportional to the $2n^{th}$ moment a $(2n+1)$-power-density is required to assure the variance of the estimator is finite. Nevertheless, this initial investigation of the IA estimator for the $n^{th}$ moment utilizes the $(n+1)$-power-density and is thus restricted in the domain of finite variance of the estimate. For the proofs, the subsamples will be assumed to be independent equals rather than approximates. First, the bias of the location and scale has the following properties.

**Lemma 4:** *a) Given $N^{(2)}$ independent-equal samples $X^{(2)}$ drawn from a 2-power Student's t-distribution $f^{(2)}\left(x\right)$ (i.e. from equal pairs of $X \sim f(x)$) the first moment, which estimates the location, is unbiased. b) Given $N^{(3)}$ samples from a centered 3-power Student's t-density $f^{(3)}\left(x\right)$, the estimate of the scale $\hat{\sigma} = 3\left(\hat{\mu}_2^{(3)} - \hat{\mu}\right)$ has a bias of $-\dfrac{1}{N^{(2)}}\sigma$.*

Proof: a) Given Lemma 3's proof that the first moment of the 2-power Student's t-density exists and is equal to the location, the estimate of the location is unbiased since

$$E\left[\frac{1}{N^{(2)}}\sum_{i=1}^{N^{(2)}}X_i^{(2)} - \mu\right] = \frac{1}{N^{(2)}}\sum_{i=1}^{N^{(2)}}E\left[X_i^{(2)}\right] - \mu = 0. \tag{5.1}$$

b) Given the unbiased estimate $\hat{\mu}$, the bias of the estimate $\hat{\sigma}$ is given by

$$E\left[\frac{3}{N^{(3)}}\sum_{i=1}^{N}\left(X_i^{(3)}-\hat{\mu}\right)^2-\sigma^2\right]=\frac{3}{N^{(3)}}\sum_{i=1}^{N}E\left[\left(X_i^{(3)}-\hat{\mu}\right)^2\right]-\sigma^2$$

$$=\frac{3}{N^{(3)}}\sum_{i=1}^{N^{(3)}}\left(E\left[\left(X_i^{(3)}-\mu\right)^2\right]-E\left[\left(\hat{\mu}-\mu\right)^2\right]\right)-\sigma^2 \qquad (5.2)$$

$$=\frac{3}{N^{(3)}}\sum_{i=1}^{N^{(3)}}\left(\frac{\sigma^2}{3}-\frac{\sigma^2}{3N^{(2)}}\right)-\sigma^2=-\frac{1}{N^{(2)}}\sigma.$$

Note that the bias of the scale estimate is inversely proportional to the number of samples used to estimate the location which is $N^{(2)}$. □

The variance properties of the estimator depend on moments twice that of the estimators $\mu_{2n}^{(n+1)}$. For the domain of (4.15), the variance of the estimator is finite for a restricted range of shape values. The variance of the location estimator depends on $\{m=2, n=2\}$ which has a finite domain of $\kappa<2$. The variance of the scale estimator depends on $\{m=4, n=3\}$ which has a finite domain of $\kappa<\frac{3}{2}$. The following lemma proves these domains and the expected value of the estimate variances.

**Lemma 5:** *a) Given $N^{(2)}$ independent-equal samples $X^{(2)}$ drawn from a 2-power Student's t-density $f^{(2)}\left(x\right)$ the first moment, which estimates the location, has a precision of*

$$\frac{\sigma}{\sqrt{\left(2-\kappa\right)N^{(2)}}}\text{ if }\kappa<2 \ . \text{ b) Given N samples from a centered 3-power Student's t-density } f^{(3)}\left(x\right),$$

*the estimate of the scale $\hat{\sigma}=3\left(\hat{\mu}_2^{(3)}-\hat{\mu}\right)$ has a precision of*

$$\frac{3\sigma^2}{\sqrt{N^{(3)}}}\left(\frac{1}{3-2\kappa}+\frac{1}{\left(\left(2-\kappa\right)N^{(2)}\right)^2}\right)^{\frac{1}{2}}\text{ if }\kappa<\frac{3}{2}.$$

*Proof: a)* The variance of the location estimate is given by

$$
\begin{aligned}
Var\left[\hat{\mu}\right] &= Var\left[\frac{1}{N^{(2)}}\sum_{i=1}^{N^{(2)}} X_i^{(2)}\right] \\
&= \frac{1}{\left(N^{(2)}\right)^2}\left[\sum_{i=1}^{N^{(2)}} Var\left[X_i^{(2)}\right]\right] \\
&= \frac{Var\left[X_i^{(2)}\right]}{N^{(2)}}
\end{aligned}
\tag{5.3}
$$

The variance of the 2-power samples from a Student's *t*-density is given by $\{m=2, n=2\}$ in (4.15)

$$
\int_{-\infty}^{\infty} x^2 f^2(x)\,dx \Big/ \int_{-\infty}^{\infty} f^2(x)\,dx = \frac{2\kappa^{-1}\sigma^2\left(\frac{1}{2}\right)!\left(\frac{2-3\kappa}{2\kappa}\right)!}{2\sqrt{\pi}\left(\frac{2-\kappa}{2\kappa}\right)!} \text{ if } \kappa < 2
\tag{5.4}
$$

$$
= \frac{\sigma^2}{2-\kappa} \text{ if } \kappa < 2.
$$

Thus, taking the square root the precision is $\dfrac{\sigma}{\sqrt{\left(2-\kappa\right)N^{(2)}}}$ if $\kappa < 2$. $\quad\square$

b) The precision of the scale estimate is given by

$$
\begin{aligned}
Var\left[\hat{\sigma}\right] &= Var\left[\frac{3}{N^{(3)}}\sum_{i=1}^{N^{(3)}}\left(X_i^{(3)}-\hat{\mu}\right)^2\right] \\
&= \frac{9}{\left(N^{(3)}\right)^2}\sum_{i=1}^{N^{(3)}}\left(Var\left[\left(X_i^{(3)}\right)^2\right]+Var\left[\hat{\mu}^2\right]\right),
\end{aligned}
\tag{5.5}
$$

since the cross-term $Var\left[2X_i^{(3)}\hat{\mu}\right]$ is zero given that the $X^{(3)}$ and $X^{(2)}$ samples are independent.

The variance of the square of the samples from a 3-power Student's *t*-density is given by $\{m=4, n=3\}$ in (4.15)

$$\int_{-\infty}^{\infty} x^4 f^3(x)dx \Big/ \int_{-\infty}^{\infty} f^3(x)dx = \frac{2\kappa^{-2}\sigma^4\left(\dfrac{3}{2}\right)!\left(\dfrac{3-4\kappa}{2\kappa}\right)!}{2\sqrt{\pi}\left(\dfrac{3}{2\kappa}\right)!} \text{ if } \kappa < \frac{3}{2}$$

(5.6)

$$= \frac{\sigma^4}{3-2\kappa}.$$

The variance of the square of the location estimate is the square of the variance of the location estimate $\dfrac{\sigma^4}{\left(\left(2-\kappa\right)N^{(2)}\right)^2}$ if $\kappa < 2$. Thus, the precision of the scale estimate is

$$\text{Prec}[\hat{\sigma}] = \sqrt{\text{Var}\left[\hat{\sigma}\right]} = \frac{3}{N^{(3)}}\left(N^{(3)}\frac{\sigma^4}{3-2\kappa} + N^{(3)}\frac{\sigma^4}{\left(\left(2-\kappa\right)N^{(2)}\right)^2}\right)^{\frac{1}{2}}$$

(5.7)

$$= \frac{3\sigma^2}{\sqrt{N^{(3)}}}\left(\frac{1}{3-2\kappa} + \frac{1}{\left(\left(2-\kappa\right)N^{(2)}\right)^2}\right)^{\frac{1}{2}} \text{ if } \kappa < \frac{3}{2}. \ \square$$

### 5.3. Performance of the IA Algorithm

The parameters of the Student's *t*-distribution are estimated using the IA algorithm by selecting approximate pairs for the location, approximate absolute value triplets for the scale, and geometric mean of all the samples to estimate the shape, which is the inverse of the degree of freedom. **Fig. 2-6** document the bias and precision of the IA estimation as the shape is increased from 0.25 to 0.4. This spans the domains of finite variance (0.25), the boundary of infinite variance (0.5), the Cauchy distribution (1) which is the focal point between a Gaussian (0) and a delta distribution ($\infty$), a stressing case (2), and approaching the limits of generating and estimating reliable distributions (4). Each figure includes a) the mean and three times the standard deviation for 20 measurements each of 25 different combinations of the location and scale, b) the mean and standard deviation for the shape and scale estimates as a function of the scale, and c) a 3D image of the bias and precision of all the measurements. In the 3D image, the location and scale measurements are scaled by σ. For the overall precision, the computation of the standard deviation over all the measurements is multiplied by $\sqrt{25}$ since it is the precision of each estimate rather than the group of estimates which is of interest. **Table 5** summarizes the bias and precision performance.

**Table 5:** The measured bias and precision for the estimates of the location, scale, and shape of the Student's *t*-distributions ranging from a shape of 0.25 to 4.00 with an initial sample size of 10,000. Up until a shape of 4, the bias for each estimate is less than 0.01 and the precision ranges from $\pm0.02$ to $\pm0.11$. The shape of 4 begins to show weaknesses but the precision is still within $\pm0.33$.

| Generated Shape | Experimental Bias and Precision | | |
|---|---|---|---|
| | Location | Scale | Shape |
| 0.25 | $0.000 \pm 0.017$ | $0.00 \pm 0.04$ | $0.00 \pm 0.07$ |
| 0.50 | $0.000 \pm 0.020$ | $0.00 \pm 0.05$ | $0.01 \pm 0.07$ |
| 1.00 | $0.001 \pm 0.024$ | $0.00 \pm 0.06$ | $0.00 \pm 0.09$ |
| 2.00 | $0.00 \pm 0.04$ | $0.00 \pm 0.10$ | $0.00 \pm 0.11$ |
| 4.00 | $0.00 \pm 0.33$ | $0.05 \pm 0.18$ | $-0.06 \pm 0.15$ |

The effect of initial sample sizes is shown in **Fig. 7** and summarized in **Table 6**. The IA algorithm is still effective with an initial sample size of 1,000 though the bias and precision are reduced relative to the 10,000 samples size. An initial sample of 100 will have cases where the subsample selection is too small for statistical estimation; however, adjustments in the algorithm may enable course estimates given a small sample size. Increasing the sample size to 100,000 shows an order of magnitude improvement (x10) in the precision over the 10,000-sample size performance when the shape is 0.25 and 1.00. For a shape of 4.00 the improvement is more modest (x2).

**Table 6:** The measured bias and precision for the estimates of the location, scale, and shape as a function of sample size and shape for the Student's *t*-distribution. For the shapes 0.25 and 1.00, the difference in precision between 1,000 and 100,000 samples is approximately a factor of 10. For shape 4.00 is only improved by a factor of 2.

| Generated Shape | Sample Size | Experimental Bias and Precision | | |
| --- | --- | --- | --- | --- |
| | | Location | Scale | Shape |
| 0.25 | 1,000 | $0.00 \pm 0.05$ | $-0.03 \pm 0.15$ | $0.04 \pm 0.19$ |
| | 100,000 | $0.000 \pm 0.006$ | $0.000 \pm 0.012$ | $0.000 \pm 0.021$ |
| 1.00 | 1,000 | $-0.02 \pm 0.07$ | $0.00 \pm 0.17$ | $0.02 \pm 0.22$ |
| | 100,000 | $0.000 \pm 0.008$ | $0.002 \pm 0.021$ | $-0.002 \pm 0.029$ |
| 4.00 | 1,000 | $0.00 \pm 0.29$ | $0.02 \pm 0.22$ | $-0.01 \pm 0.21$ |
| | 100,000 | $0.00 \pm 0.14$ | $0.05 \pm 0.11$ | $-0.04 \pm 0.09$ |

**Table 7** shows the number of pairs and triplets selected as a function of the initial sample size and the shape of the distribution. Each estimate is performed by the selection of pairs to estimate the location and triplets to estimate the scale. To ensure an adequate number of samples, a type of bootstrapping is utilized in which the original sample is permutated 10 times. The pairs and triplets are partitioned without an offset, which does create some correlation in the selection due to the overlap of groups; however, the alternative is to complete more permutations which is the slowest part of the algorithm. From the mean pairs and triplets selected the theoretical bias and precision are determined using Lemma 4 and Lemma 5, respectively. For the shape of 0.25 and 1.00, the theoretical and experimental bias and precision are similar. For a shape of 4, the precision is not finite, which is not fully reflected in the experimental measurements but could be shown by further increases in the sample size. A follow-up study is planned to examine the optimization of the speed and accuracy of the algorithm, which will examine more closely the effect of the tolerance, partition offset, and permutations.

**Table 7:** Relationship between original sample size, selection of pairs and triplets, and the bias and precision for the Student's *t* estimates. The mean and standard deviation of the pairs and triplets selected are computed over the 20 estimates and 25 different location and scale values. Each selection includes 10 permutations of the initial samples. The bias and precision of the location and scale are determined from Lemma 4 and Lemma 5, respectively using the mean pair and triplet values.

| Generated Shape | Sample Size | Pairs Selected | Triplets Selected | Theoretical Bias $\pm$ Precision | |
| --- | --- | --- | --- | --- | --- |
| | | | | Location | Scale |
| 0.25 | 1,000 | $9 \pm 1$ | $7 \pm 1$ | $0.00 \pm 0.25$ | $-0.1 \pm 0.7$ |
| | 10,000 | $90 \pm 3$ | $71 \pm 4$ | $0.00 \pm 0.08$ | $-0.01 \pm 0.23$ |
| | 100,000 | $893 \pm 8$ | $709 \pm 13$ | $0.000 \pm 0.025$ | $0.00 \pm 0.07$ |
| 1.00 | 1,000 | $8 \pm 1$ | $7 \pm 2$ | $0.00 \pm 0.35$ | $-0.1 \pm 1.2$ |
| | 10,000 | $79 \pm 2$ | $68 \pm 4$ | $0.00 \pm 0.11$ | $0.0 \pm 0.4$ |
| | 100,000 | $794 \pm 7$ | $681 \pm 14$ | $0.000 \pm 0.035$ | $0.00 \pm 0.12$ |
| 4.00 | 1,000 | $11 \pm 2$ | $24 \pm 8$ | $0 \pm \infty$ | $0 \pm \infty$ |
| | 10,000 | $118 \pm 7$ | $261 \pm 38$ | $0 \pm \infty$ | $0 \pm \infty$ |
| | 100,000 | $1170 \pm 26$ | $2605 \pm 128$ | $0 \pm \infty$ | $0 \pm \infty$ |

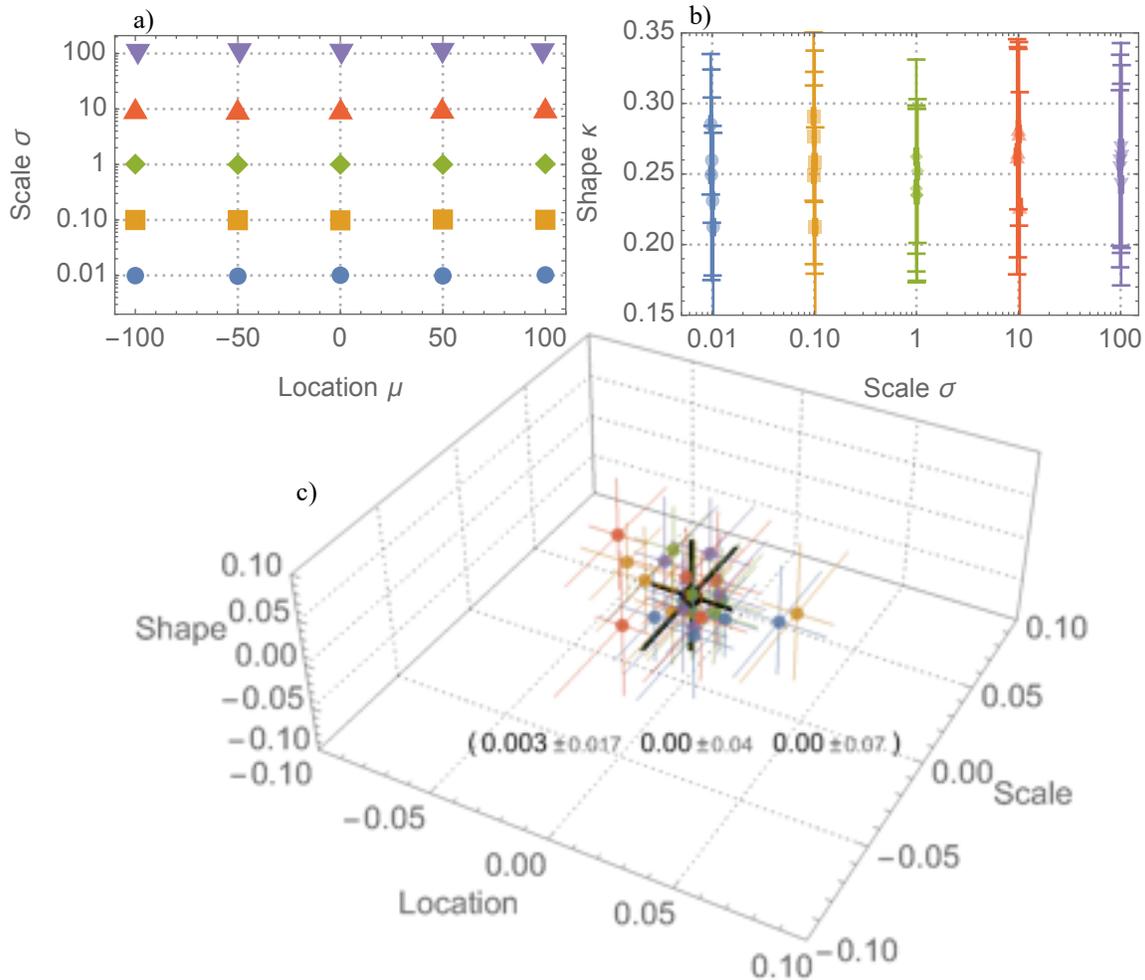

**Fig. 2. Student's *t* (Shape = 0.25) Estimation Performance** Estimation performance for the location, scale, and shape (0.25) Student's t distribution with 10,000 initial samples. a) Location (mean) versus the scale for 25 different estimates each performed 20 times. Three times the standard deviation are shown as error bars. b) Scale versus the shape. The error bars are the standard deviation. b) Scale versus the shape. The error bars are the standard deviation of 20 estimates.  c) Bias (estimate minus truth) and precision (standard deviation) for the three parameter estimations. The colors indicate the scale as in a and b. The location and scale error is divided by the scale, while the coupling error is not scaled. The average performance over all the results is in black and also as a text inset (location, scale, shape).

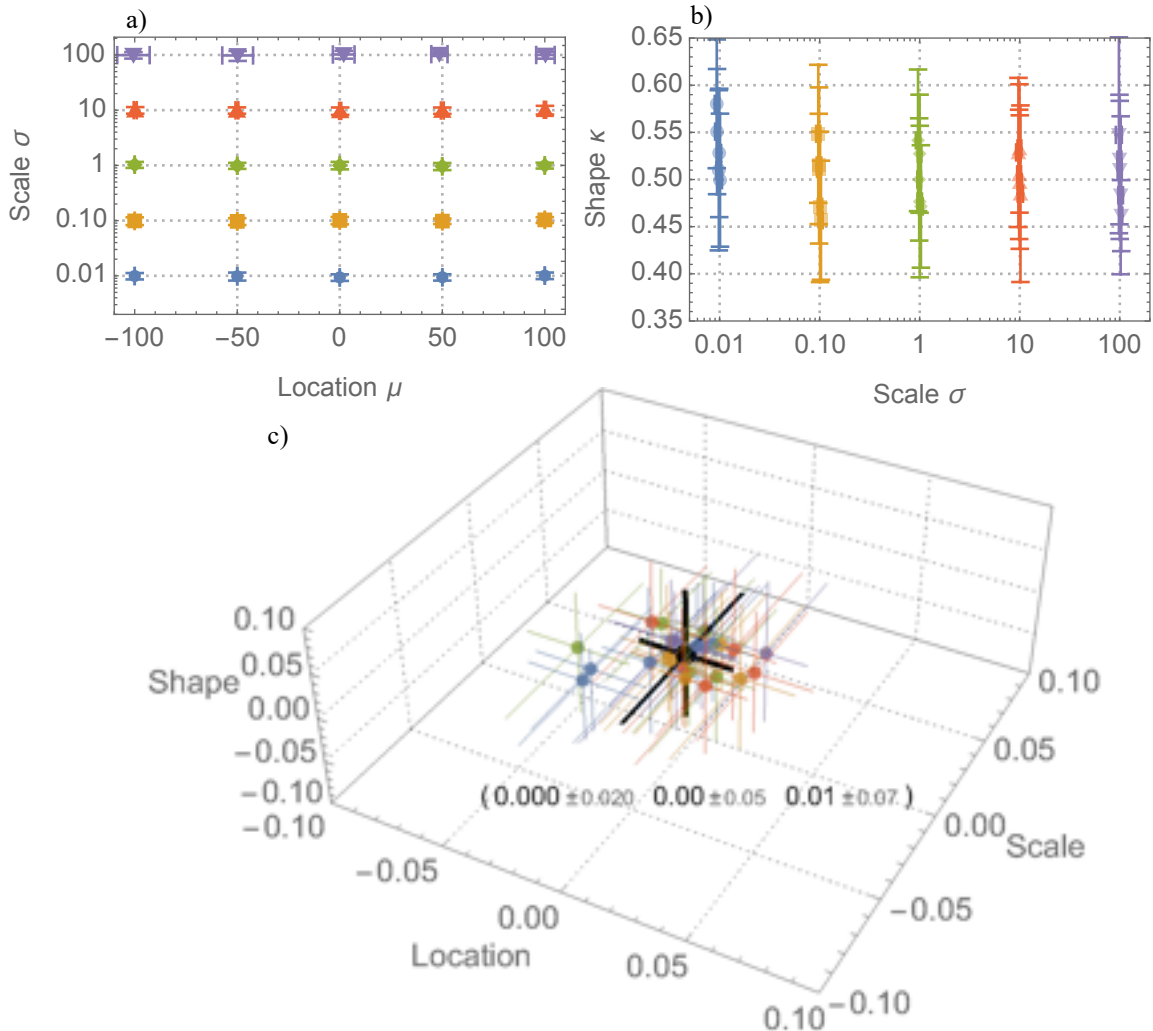

**Fig. 3. Student's *t* (Shape = 0.5) Estimation Performance** Estimation performance for the location, scale, and shape (0.50) Student's t distribution with 10,000 initial samples. a) Location (mean) versus the scale for 25 different estimates each performed 20 times. Three times the standard deviation are shown as error bars. b) Scale versus the shape. The error bars are the standard deviation. c) Bias (estimate minus truth) and precision (standard deviation) for the three parameter estimations. The colors indicate the scale as in a and b. The location and scale error is divided by the scale, while the coupling error is not scaled. The average performance over all the results is in black and also as a text inset (location, scale, shape).

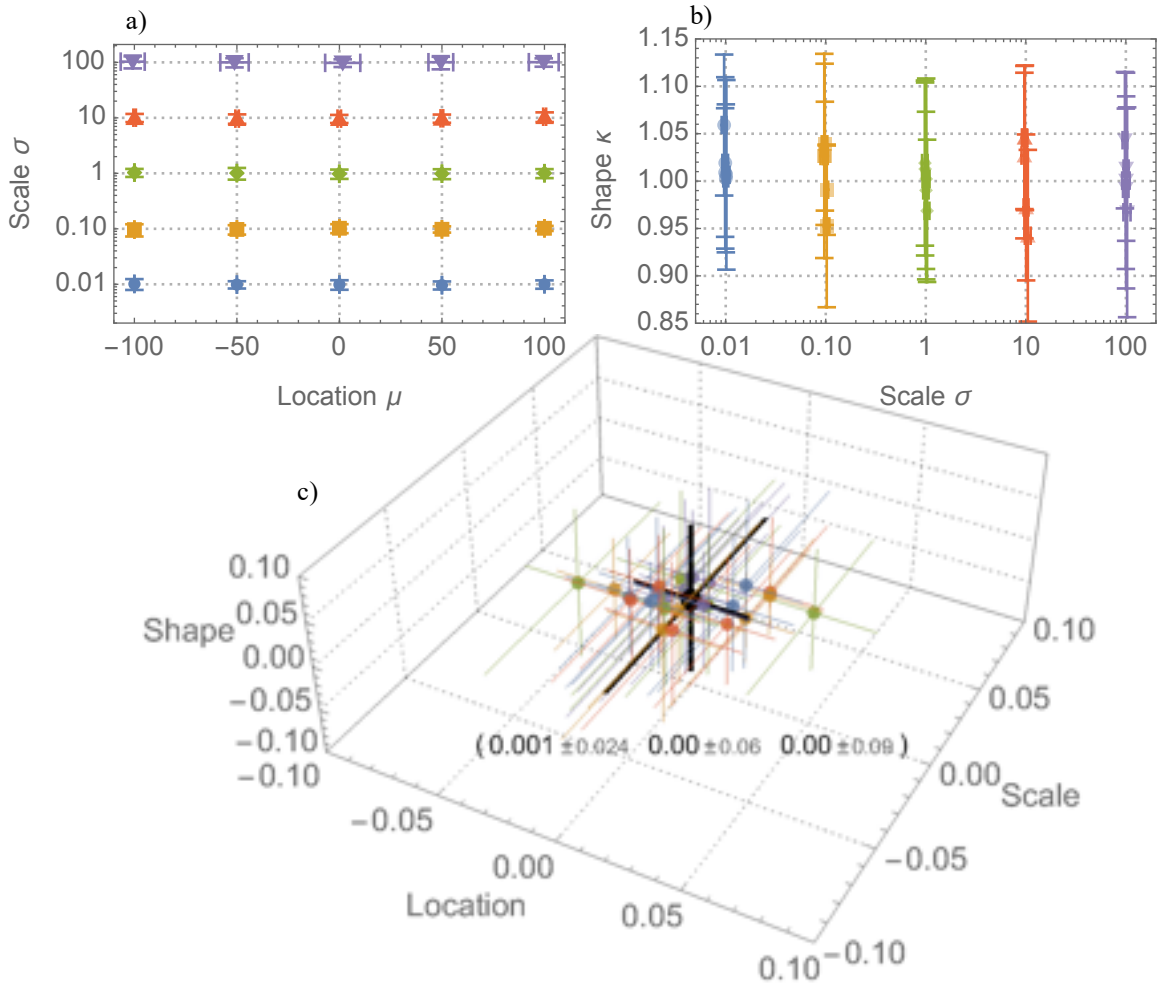

**Fig. 4. Student's *t* (Shape = 1.0) Estimation Performance** Estimation performance for the location, scale, and shape (1.0) Student's t distribution with 10,000 initial samples. a) Location (mean) versus the scale for 25 different estimates each performed 20 times. Three times the standard deviation are shown as error bars. b) Scale versus the shape. The error bars are the standard deviation. c) Bias (estimate minus truth) and precision (standard deviation) for the three parameter estimations. The colors indicate the scale as in a and b. The location and scale error is divided by the scale, while the coupling error is not scaled. The average performance over all the results is in black and also as a text inset (location, scale, shape).

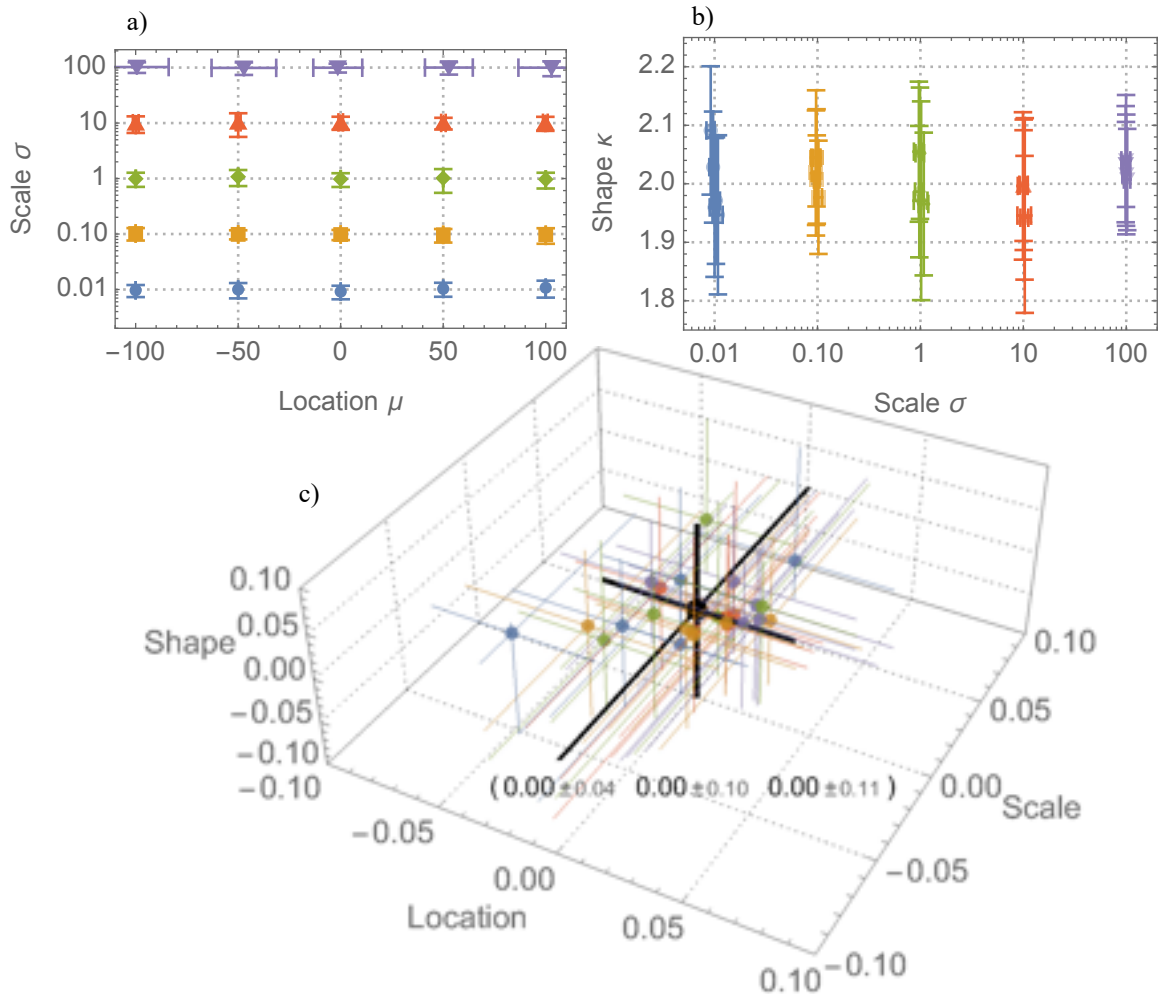

**Fig. 5. Student's *t* (Shape = 2.0) Estimation Performance** Estimation performance for the location, scale, and shape (2.0) Student's t distribution with 10,000 initial samples. a) Location (mean) versus the scale for 25 different estimates each performed 20 times. Three times the standard deviation are shown as error bars. b) Scale versus the shape. The error bars are the standard deviation.  c) Bias (estimate minus truth) and precision (standard deviation) for the three parameter estimations. The colors indicate the scale as in a and b. The location and scale error is divided by the scale, while the coupling error is not scaled. The average performance over all the results is in black and also as a text inset (location, scale, shape).

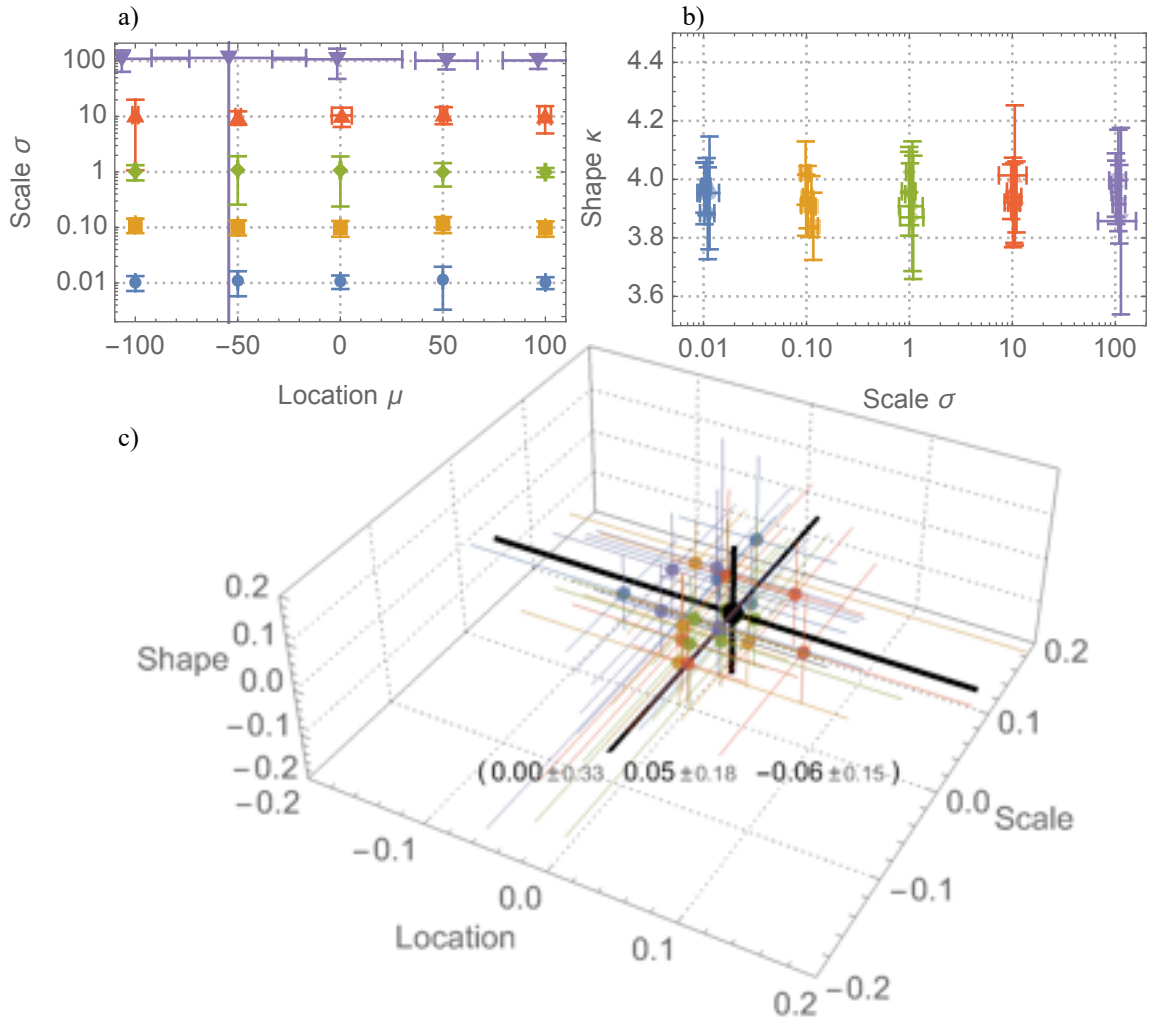

**Fig. 6. Student's *t* (Shape = 4) Estimation Performance** Estimation performance for the location, scale, and shape (4.0) Student's t distribution with 10,000 initial samples. a) Location (mean) versus the scale for 25 different estimates each performed 20 times. Three times the standard deviation are shown as error bars. b) Scale versus the shape. The error bars are the standard deviation. c) Bias and precision which together quantify the accuracy of the estimator. The colors indicate scale. The location and scale error is divided by the scale, while the coupling error is not scaled. The average performance over all the results is in black and is inset as (location, scale, shape).

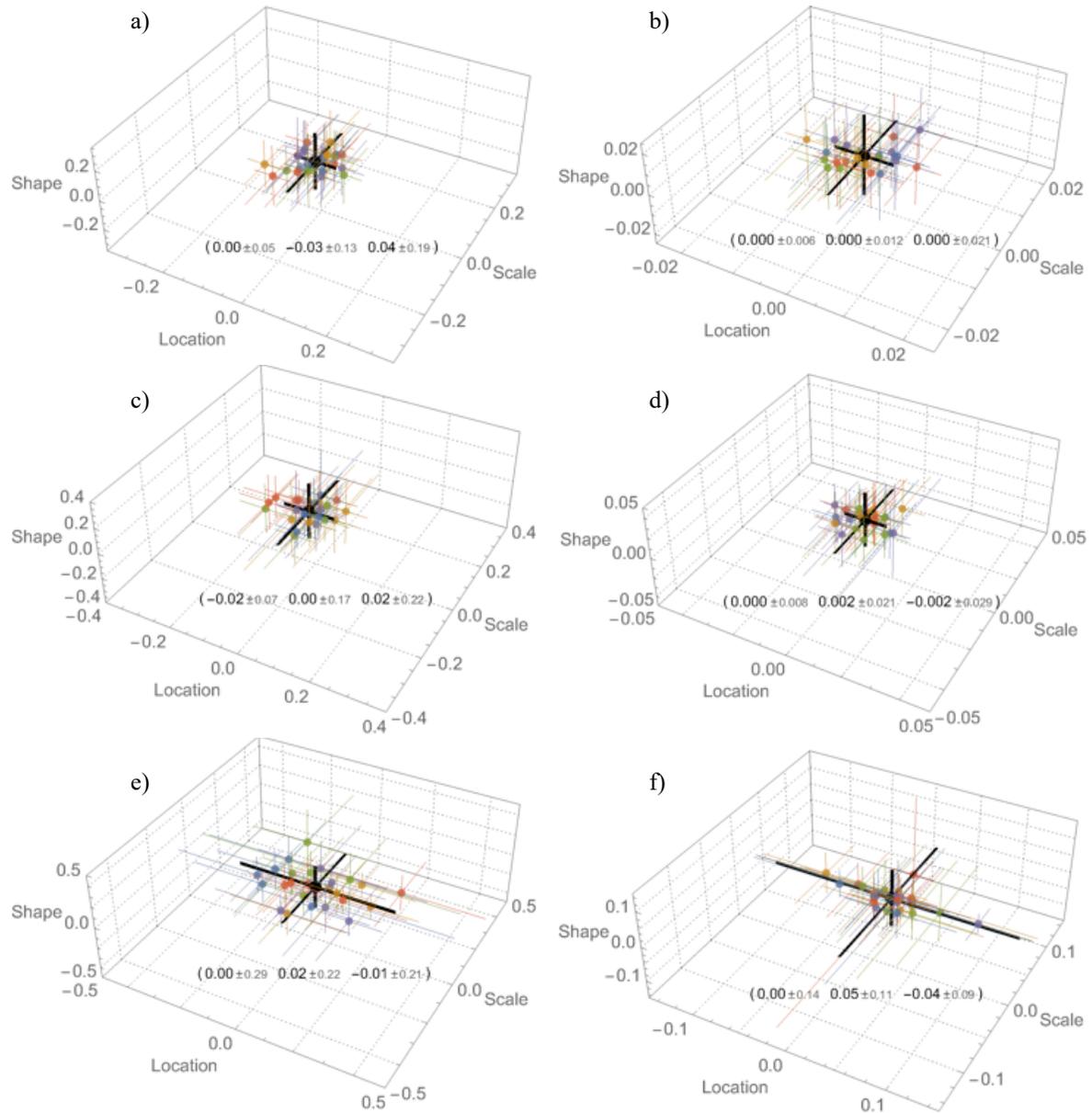

**Fig. 7. Student's *t* estimation performance for 1000 and 100,000 samples.** The bias and precision of mean, scale, and shape for a sample size of 1,000 (left) and 100,000 (right). The rows from top to bottom have shape values of 025, 1.0, and 4.0. The 100-fold increase in the sample size improves the precision of the estimate by approximately a factor of 10 for shape 0.25 and 1.0 and a factor 2 for the shape of 4.0.

## 6.  Comparative Performance and Applications

### 6.1. Small-sample estimation performance

Many applications require estimation with small datasets. For instance, astronomical data is often sparse. As an example, the solar wind, a stream of plasma emanating from the sun, has been estimated to have a Student's t distribution[2] based on datasets of less than 1000 samples. For this reason, we take a closer look at the accuracy and sensitivity of the IA algorithm with a sample size of 100 and then apply the findings to the processing of solar wind data. The performance is compared with the maximum likelihood estimate using two metrics: the negative loglikelihood and the Cramer-von Mises p-value. The stability of the loglikelihood measurement is helpful given the heavy-tail estimates being evaluated. The Cramer-von Mises p-value also provides insight but its basis as a mean-square error evaluation can be highly sensitive to samples in the tails.

The IA algorithm has two control parameters, the number of permutations for grouping samples and the domain for selecting groups that are approximately equal. The permutations are regroupings of the pairs and triplets, and thus increase selections at the expense of adding some correlation. The selection domain is relative to the current estimate of the scale (the 75% percentile for the first selection), and the trade-off is between increasing the samples and deviating from the distribution along the diagonal. **Fig. 9** shows the mean and standard deviation of the negative loglikelihood and Cramer-von Mises p-value for a variety of IA estimates and the maximum likelihood estimate. The results are based on 25 estimates with 100 samples. Three permutations (1, 5, and 10) and three selection domains (0.5, 1, and 2) are evaluated. While just one permutation has the risk of inadequate samples for an estimate, this is eliminated by increasing the domain selection to 2. Performance equal to or very close to the maximum likelihood estimate is achieved for all four combinations of permutations of 5 or 10 and domain selection of 0.5 or 1. Thus even for small sample sizes the algorithm shows robustness across a range of parameter settings.

### 6.2. Astrophysics data of the solar wind

Burlaga and Ness (2013) reported on the distribution of daily averages and standard deviations of magnetic fields of the heliosphere solar wind, which are plasma streams from the sun. Their report, specified in terms of $q$-Gaussians and translated here to scale and shape, included 331 days of the average magnetic field for the year 2010, the daily incremental changes, and the daily incremental changes in the standard deviation (Ness and Richardson 2020). **Fig. 8** shows the daily time series for the magnetic field strength and its increments with standard deviation error bars. As shown in **Fig. 10** the Independent Approximates estimations closely match the Maximum Likelihood estimation, as computed using the *Mathematica* FindDistributionParameters function. The IA pair and triplet selection settings were set to a domain of one (relative to the scale) and just one permutation. The average of 25 estimations is shown.  The loglikelihood of the IA estimates are within 0.5% of the maximum likelihood estimate.  The performance metrics of the Burlaga and Ness estimates are not shown since there are reasons for the differences. The magnitude distribution assumed a Gaussian, and there appears to be some slight differences in the incremental changes' datasets.

---

[2] The astrophysical community sometimes uses the kappa-distribution which, along with $q$-Gaussian and coupled-Gaussian, is equivalent to the Student's t distribution.

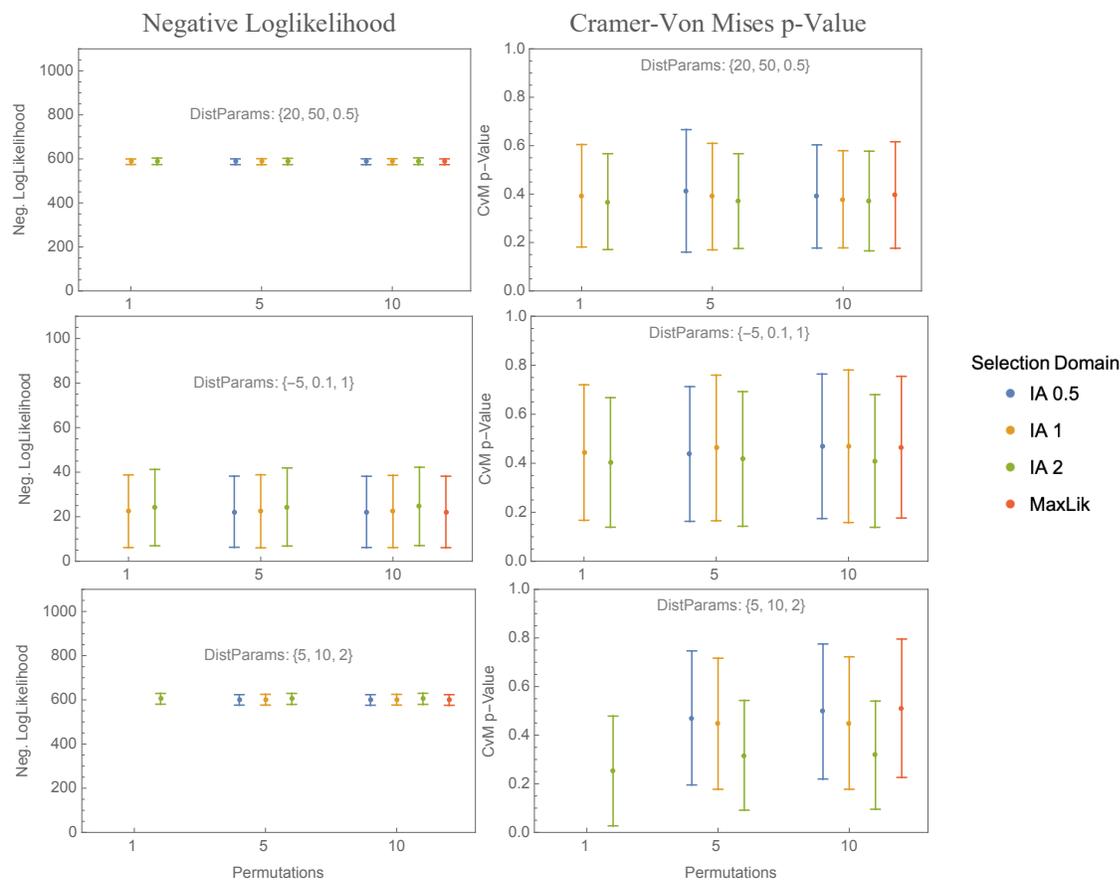

**Fig. 9. Comparative Performance with 100 samples.** The negative loglikelihood and the Cramer-von Mises p-values given 100 samples for the IA estimation is comparable with the maximum likelihood estimate across a variety of selection domains and permutations. Standard deviation error bars are shown for 25 estimates. For one permutation there is a risk no IA samples, so the smaller selection domains 0.5 (blue) and 1 (orange) are missing. With the larger selection domain of 2 (green) just one permutation is adequate to achieve an estimate. All of the IA estimates are within the standard deviation error of the maximum likelihood estimate (red). IA estimate performance at or very close to the maximum likelihood performance is achieved for all combinations of 5 or 10 permutations and 0.5 or 1 selection domain.

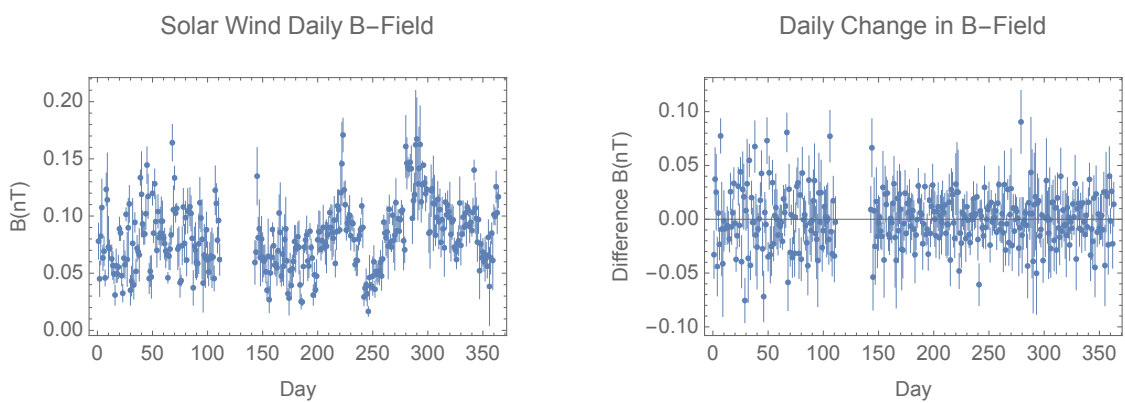

**Fig. 8. Solar Wind Data.** a) The daily average magnetic field strength with standard deviation of the heliosphere solar wind for 2010. b) The daily incremental change in the magnetic field.

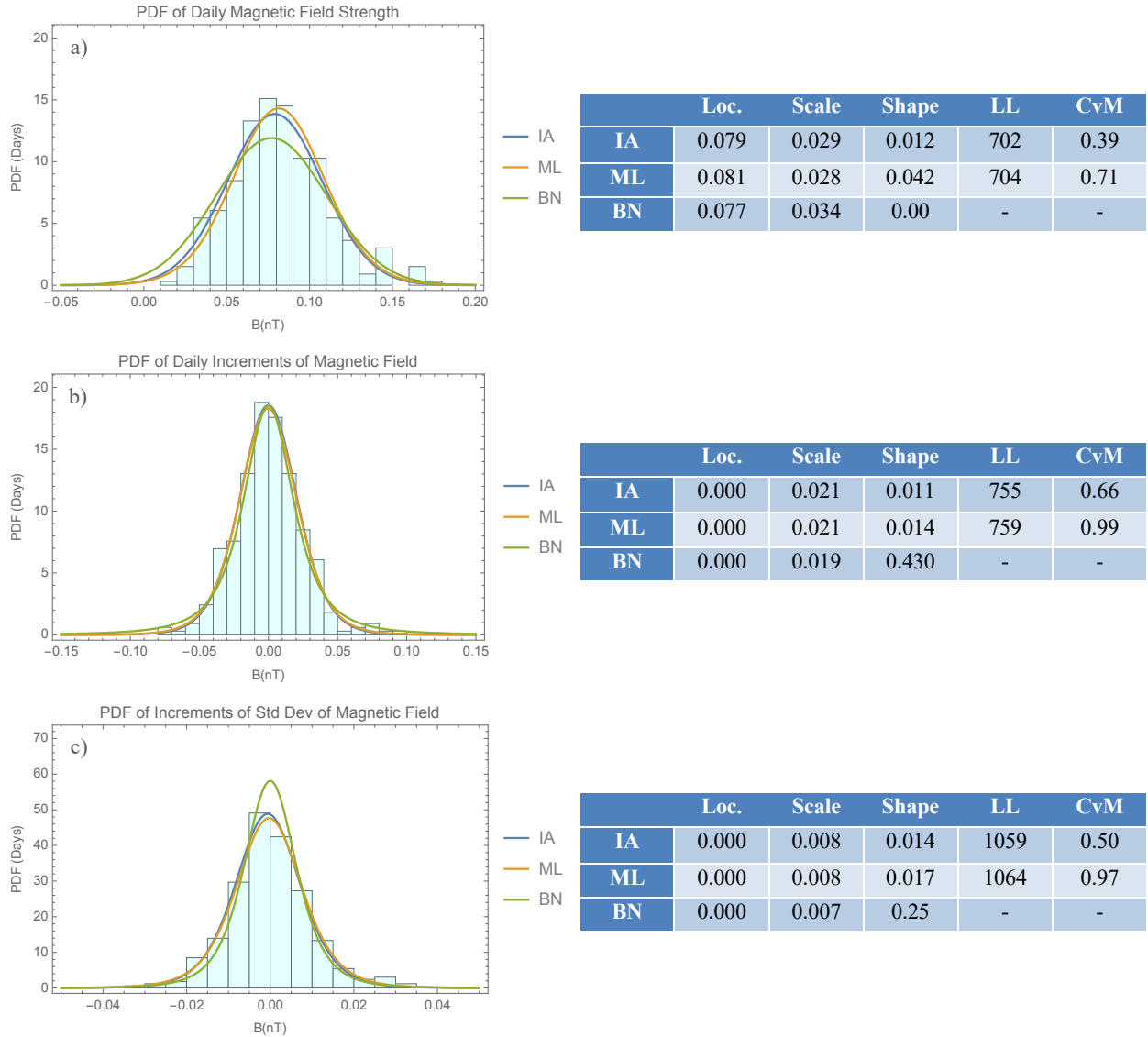

**Fig. 10. Solar Wind Distributions.** The histogram, estimated pdf, and table of parameters and performance are shown for a) the daily magnetic field strength, b) the daily incremental change in the magnetic field strength, and c) the daily incremental change in the standard deviation of the magnetic field. Each plot and table show the estimations for Independent Approximates algorithm (IA), the maximum likelihood (ML), and the report in (Burlaga and Ness, 2013).

## 6.3. Large sample simulation of the Standard Map

Nonlinear systems, even if deterministic, can produce a variety of complex patterns that require statistical analysis. An example of this is the Standard Map, which is governed by an iterative function of two variables:

$$x_{n+1} = x_n + y_{n+1}; \;\; y_{n+1} = y_n + \text{K} \, sin(x_n) \tag{6.1}$$

The parameter $K$ controls the degree of nonlinearity introduced into the sequence, with $K = 0$ linear and $K = 10$ so chaotic that the map produces states distributed as a Gaussian distribution. In (Tirnakli and Borges 2015; Tirnakli and Tsallis 2020) the statistics of the Standard Map for a large number of initial conditions ($M \geq 10^7$) are reported on. They show that in the linear region, $K = 0.0$, the statistics are heavy-tailed and approach a Cauchy distribution. Intermediate values of $K$ have a mixture of chaotic Gaussian regions and complex near-Cauchy regions. The statistics are measured by summing over the T iterative values of $x$ and subtracting the average,

$$z := \sum_{i=1}^{T}(x_i - \langle x \rangle)$$
$$\langle x \rangle = \frac{1}{M} \frac{1}{T} \sum_{j=1}^{M} \sum_{i=1}^{T} x_i^{(j)}. \tag{6.2}$$

The random variable $z$ is thus subject to properties of the central limit theorem. For 200 million samples with $K = 0$ Tirnakli used a graphical visualization to estimate $q = 1.935$ which determines the shape according to (2.10) and $f(0) = 3.30$ which determines the scale via the normalization term of (2.5).[3]

    For the dataset of 200 million samples with $K = 0$ the Independent Approximate estimate is shown to be consistent with a maximum likelihood (ML) estimate using the *Mathematica* EstimatedDistribution function and a Hill estimate of the tail shape combined with ML estimates of the location and scale. The Hill estimator was estimated by computing the shape value over a broad range of the $k$-highest order parameters, and then averaging over a stable subset. Table 8 summarizes the average loglikelihood and the Cramer-von Mises (CvM) p-values of the methods for 200 million samples. The three analytical estimates have an identical average loglikelihood of 0.114, while the CvM p-value of the IA method is slightly better (86.5 x $10^{-15}$). The Tirnakli visualization estimated has a lower average loglikelihood (0.103) but a higher CvM p-value (320 x $10^{-15}$).

    **Fig. 11** shows a log-log plot of the data histogram and the estimated distributions. The four estimates are visually consistent with each other. There is a knee near the scale of approximately 0.08. Below the knee, the histogram has some deviations from the tail region of a Student's t distribution. Above the knee, there is heavy-tail decay consistent with estimates of a shape of approximately 0.9. Table 8 also shows the evaluation of the models in just the middle domain from $10^{-3}$ to $10^3$. In this case the IA method shows a slight improvement over the ML and Hill estimates in both average loglikelihood and the CvM metrics, while the Tirnakli estimate now shows lower values in both metrics. The relatively high CvM value with all the data for the Tirnakli estimate is likely due to the extreme values, where the mean-square average of the CvM is more sensitive.

---

[3] In (Tirnakli and Tsallis 2020) $f(0)$ is reported to be 0.0363 for $K = 0$; however, this was an error and the estimate was revised to 3.30 for this analysis. The original estimate of $q = 1.935$ is unchanged and equals the IA estimate.

**Table 8.** Comparison of parameter estimation for Student's t for 200 million samples from a Standard Map set for linear dynamics ($K = 0$). The IA, Maximum Likelihood and Hill estimate each have an average loglikelihood of 0.114. The CvM p-value of the IA method is slightly higher (86.5 x $10^{-15}$) than ML and Hill estimates. The Tirnakli report accurately estimates the shape but overestimates the scale. The mismatch is reflected in a much lower average loglikelihood (-3.33) but the CvM p-value is actually higher (4.02 x $10^{-12}$). Examination of just the middle domain (0.001 to 1000) results in a relative reduction in the CvM p-value of the Tirnakli report.

| Method | Location | Scale | Shape | Avg LL | CvM p-value | Avg LL Middle | CvM Middle |
|---|---|---|---|---|---|---|---|
| **Indep. Approx.** | 5.14 x $10^{-4}$ | 0.0775 | 0.878 | 0.114 | 86.5 x $10^{-15}$ | 0.0391 | 3.01 x $10^{-12}$ |
| **Max Likelihood** | 9.34 x $10^{-6}$ | 0.0800 | 0.893 | 0.114 | 57.5 x $10^{-15}$ | 0.0378 | 2.98 x $10^{-12}$ |
| **Hill & ML** | 9.35 x $10^{-6}$ | 0.0790 | 0.926 | 0.114 | 54.3 x $10^{-15}$ | 0.0375 | 2.98 x $10^{-12}$ |
| **Tirnakli Visual** | 0.00 | 0.0989 | 0.878 | 0.103 | 320 x $10^{-15}$ | 0.0351 | 2.78 x $10^{-12}$ |

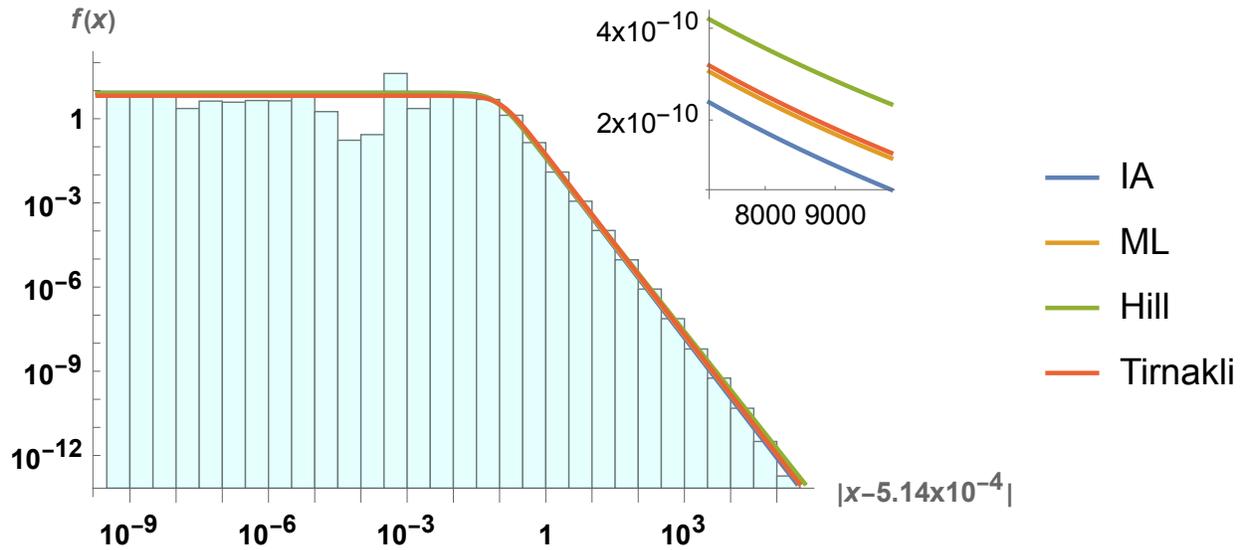

**Fig. 11.** Histogram (bars) and model fits (lines) for 200 million samples from a Standard Map simulation in the linear domain with $K = 0$. The four estimates, Independent Approximates, Maximum Likelihood (ML), Hill, and the Tirnakli, are in close agreement.

## 7. Conclusion and Future Research

A closed-formed method for estimating heavy-tailed distributions has been discovered that relies on a filtering method that selects *Independent Approximates*. The IAs are selected by partitioning independent, identically distributed samples into partitions of size *n* and selecting the median of only the groups approximately equal within a tolerance. The IAs are approximately distributed as the normalized power-density $f_X^n(x) \Big/ \int_{x \in X} f_X^n(x)dx$. Because the power-density preserves a precise relationship to properties of *f* while increasing the rate of decay of the tail, estimates of the parameters of *f* are enabled. The discovery clarifies that the *q* parameter of nonextensive statistical mechanics or *q*-statistics quantifies the number of random variables which are equal. For integer values then, the power-density referred to as an escort probability in *q*-statistics is the density of a subsample of IAs. Further research is suggested to specify the subsampling with a fractional power-density. The fractional power-density is conjectured to be either a non-equal marginal or a fractal representation of partitions.

To the author's knowledge, having completed a thorough literature search, this is the first report of such a filtering method to estimate heavy-tailed distributions. A proof applicable to distributions with an analytical density is completed for the Type II Pareto and Student's *t*. Numerical experiments for the Student's *t* distribution demonstrate that estimates of the location, scale, and shape with relative biases below 0.01 and precision below $\pm 0.1$ are achieved with an initial sample size of 10,000. The estimates approximate the precision of a maximum likelihood search, even with small sample sizes, such as 100. For large sample sizes, the closed-form computation reduces the required computations.

The technique opens many new avenues for additional research. The use of quartets of IAs is expected to enable estimation of the heavy-tailed distributions with skew such as the Pareto Type IV. Proofs regarding the performance of the algorithm as a function of sample size, approximation tolerance, and the number of permutations used in the selection process would be an important contribution to the research. Given that even in the small size cases, accurate estimates were achieved without permutation sampling (a slow but perhaps unnecessary part of the algorithm), comparative measurements of the IA algorithm speed would be illuminating. Use of groupings of 4 or higher lower the theoretical variance of the estimates of the location and scale but at the expense of reduced subsamples. The cost-benefit analysis of this trade-off would be an interesting study.

A variety of applications are anticipated as industrial processes still rely heavily on estimates of the mean and standard deviation even when the underlying distributions are non-Gaussian. For instance, in the financial sector, events such as the 2008 financial crisis are understood to have been exacerbated by methods such as the Black-Sholes options pricing (Goldstein and Taleb 2007; Black and Scholes) that relied on assumptions about independence and Gaussian processes. The Volatility Index (VIX) underlying pricing of many financial derivatives is closely related to the log-return standard deviation despite long-standing evidence that the variations in log-returns are heavy-tailed and thus not characterized by the standard deviation (Kapadia and Du 2011; Park et al. 2017; Zubillaga et al. 2022).

In turn, scientific investigations of complex systems have relied on a variety of iterative methods that are cumbersome to describe and implement. In medicine, vital signs such as heartbeats are known to follow fractal rhythms (Goldberger et al. 2002; Selvaraj et al. 2011); fast, accurate measures of the distributions of vital signs could improve diagnostics. In addition to estimation, signal processing applications particularly in digital media and aerospace instrumentation may be able to use the IA method to filter outliers while preserving core signal information.

## ACKNOWLEDGMENTS

Discussions with William Thistleton and Mark Kon were helpful in reviewing the manuscript. Sabir Umarov's mentorship in defining the principles of *nonlinear statistical coupling* was instrumental in developing new perspectives on statistical estimation. Amenah Al-Najafi has begun studies of the generalized Pareto estimation. Ugur Tirnakli and Leonard Burlaga provided guidance on the statistical physics datasets.

## REFERENCES

Amari SI, Ohara A, Matsuzoe H. 2012. Geometry of deformed exponential families: Invariant, dually-flat and conformal geometries. Phys A Stat Mech its Appl. 391(18):4308–4319. doi:10.1016/j.physa.2012.04.016. http://dx.doi.org/10.1016/j.physa.2012.04.016.

Aschwanden MJ. 2011. Self-organized criticality in astrophysics : the statistics of nonlinear processes in the universe. Springer Science & Business Media. [accessed 2020 Oct 24]. https://ui.adsabs.harvard.edu/abs/2011soca.book.....A/abstract.

Beck C. 2009. Generalised information and entropy measures in physics. Contemp Phys. 50(4):495–510.

Black F, Scholes M. The Pricing of Options and Corporate Liabilities. [accessed 2020 Nov 26]. https://www-jstor-org.ezproxy.bu.edu/stable/pdf/1831029.pdf?refreqid=excelsior%3A8cc4597e3ec25bf29d4cea8659ad1b9c.

Burlaga LF, Ness NF. 2013. MAGNETIC FIELD STRENGTH FLUCTUATIONS AND THE $q$ -TRIPLET IN THE HELIOSHEATH: *VOYAGER 2* OBSERVATIONS FROM 91.0 TO 94.2 AU AT LATITUDE 30° S. Astrophys J. 765(1):35. doi:10.1088/0004-637X/765/1/35. http://stacks.iop.org/0004-637X/765/i=1/a=35?key=crossref.6b40b11079a80a95dba5c292532488ee.

Cirillo R. 2012. The Economics of Vilfredo Pareto. Routledge. [accessed 2020 Oct 24]. https://www.taylorfrancis.com/books/9780203061466.

Clauset A, Shalizi CR, Newman MEJ. 2009. Power-Law Distributions in Empirical Data. SIAM Rev. 51(4):661–703. doi:10.1137/070710111. [accessed 2020 Dec 9]. http://epubs.siam.org/doi/10.1137/070710111.

Fedotenkov I. 2018. A review of more than one hundred Pareto-tail index estimators. Munich Personal RePEc Archive.

Ferrari D, Yang Y. 2010. Maximum Lq-likelihood estimation. Ann Stat. 38(2):753–783.

Gayen A, Kumar MA. 2018. Generalized Estimating Equation for the Student-t Distributions. In: 2018 IEEE International Symposium on Information Theory (ISIT). Vail, CO. p. 1–6. [accessed 2018 May 9]. https://arxiv.org/pdf/1801.09100v1.pdf.

Goldberger AL, N Amaral LA, Hausdorff JM, Ch Ivanov P, Peng C, Stanley HE. 2002. Fractal dynamics in physiology: Alterations with disease and aging. Proc NatL Acad Sci USA. 99(1):2466–2472. [accessed 2020 Nov 26]. www.pnas.orgcgidoi10.1073pnas.012579499.

Goldstein DG, Taleb NN. 2007. We Don't Quite Know What We Are Talking About. J Portf Manag.

33(4):84–86. doi:10.3905/jpm.2007.690609. [accessed 2020 Nov 26]. http://jpm.pm-research.com/lookup/doi/10.3905/jpm.2007.690609.

Gossett W. 1904. The Application of the "Law of Error" to the Work of the Brewery. Dublin.

Hanel R, Corominas-Murtra B, Liu B, Thurner S. 2017. Fitting power-laws in empirical data with estimators that work for all exponents. Altmann EG, editor. PLoS One. 12(2):e0170920. doi:10.1371/journal.pone.0170920. [accessed 2020 Dec 2]. https://dx.plos.org/10.1371/journal.pone.0170920.

Hill BM. 1975. A simple general approach to inference about the tail of a distribution. Ann Stat. 3(5):1163–1174.

Kapadia N, Du J. 2011. The Tail in the Volatility Index. In: Fifth Singapore International Conference on Finance. [accessed 2020 Nov 26]. http://www.ssrn.com/abstract=1746528.

Katz RW, Brush GS, Parlange MB. 2005. Statistics of Extremes: Modeling Ecological Disturbances. Ecology. 86(5):1124–1134. doi:10.1890/04-0606. [accessed 2020 Oct 24]. http://doi.wiley.com/10.1890/04-0606.

Kotz S, Nadarajah S. 2004. Multivariate T-Distributions and Their Applications. Cambridge University Press. [accessed 2020 Oct 24]. https://ideas.repec.org/b/cup/cbooks/9780521826549.html.

Lévy P. 1948. Processus stochastiques et mouvement brownien. Paris: Gautheir-Villars. [accessed 2022 Feb 8]. https://scholar.google.com/scholar?hl=en&as_sdt=0%2C22&q=1948+-+Processus+stochastiques+et+mouvement+brownien+levy&btnG=#d=gs_cit&u=%2Fscholar%3Fq%3Dinfo%3A1UfA6z_R3ZsJ%3Ascholar.google.com%2F%26output%3Dcite%26scirp%3D0%26hl%3Den.

López-Ruiz R, Mancini HL, Calbet X. 1995. A statistical measure of complexity. Phys Lett A. 209(5–6):321–326. doi:10.1016/0375-9601(95)00867-5.

Nelson K, Thistleton W. 2006 May 23. Addendum: Generalized Box-Muller method for generating q-Gaussian random deviates. arXiv:0605570v2 [cond-mat]. [accessed 2020 Dec 4]. http://arxiv.org/abs/cond-mat/0605570.

Nelson KP. 2015. A definition of the coupled-product for multivariate coupled-exponentials. Phys A Stat Mech its Appl. 422:187–192. doi:10.1016/j.physa.2014.12.023. [accessed 2016 Jan 29]. http://www.sciencedirect.com/science/article/pii/S0378437114010589.

Nelson KP. 2020a. Independent-Approximates. Github. https://github.com/Photrek/Independent-Approximates.

Nelson KP. 2020b. Nonlinear-Statistical-Coupling. Github. https://github.com/kenricnelson/Nonlinear-Statistical-Coupling.

Nelson KP, Kon MA, Umarov SR. 2019. Use of the geometric mean as a statistic for the scale of the coupled Gaussian distributions. Physica A. 515:248–257. doi:10.1016/j.physa.2018.09.049. https://doi.org/10.1016/j.physa.2018.09.049.

Nelson KP, Umarov S. 2010. Nonlinear statistical coupling. Phys A Stat Mech its Appl. 389(11):2157–2163. doi:10.1016/J.PHYSA.2010.01.044. [accessed 2019 May 14]. https://www.sciencedirect.com/science/article/pii/S0378437110000993.

Nelson KP, Umarov SR, Kon MA. 2017. On the average uncertainty for systems with nonlinear coupling. Phys A Stat Mech its Appl. 468:30–43. doi:10.1016/j.physa.2016.09.046.

http://dx.doi.org/10.1016/j.physa.2016.09.046.

Ness NF, Richardson J. 2020. NASA CDAWeb DATA Explorer. VOYAGER2_COHO1HR_MERGED_MAG_PLASMA. [accessed 2022 Mar 23]. https://cdaweb.gsfc.nasa.gov/cgi-bin/eval2.cgi?dataset=VOYAGER2_COHO1HR_MERGED_MAG_PLASMA&index=sp_phys.

Nielsen F. 2013. Cramer-Rao Lower bound and Information Geometry. In: Connected at Infinity II. Gurgaon: Hindustan Book Agency.

Pareto V. 1896. Cours d'Economie Politique: professé à l'Universjté de Lausanne. F. Rouge. [accessed 2022 Feb 8]. https://zenodo.org/record/2144014.

Park S-K, Choi J-E, Shin DW. 2017. Value at risk forecasting for volatility index. Appl Econ Lett. 24(21):1613–1620. doi:10.1080/13504851.2017.1366631. [accessed 2020 Nov 26]. https://www.tandfonline.com/doi/full/10.1080/13504851.2017.1366631.

Piantadosi ST. 2014. Zipf's word frequency law in natural language: A critical review and future directions. Psychon Bull Rev. 21(5):1112–1130. doi:10.3758/s13423-014-0585-6. [accessed 2020 Oct 24]. http://link.springer.com/10.3758/s13423-014-0585-6.

Pisarenko V, Sornette D. 2006. New statistic for financial return distributions: Power-law or exponential? Phys A Stat Mech its Appl. 366:387–400. doi:10.1016/J.PHYSA.2005.10.015. [accessed 2020 Dec 9]. https://www.sciencedirect.com/science/article/pii/S0378437105010885.

Qin Y, Priebe CE. 2013. Maximum Lq-likelihood estimation via the expectation-maximization algorithm: A robust estimation of mixture models. J Am Stat Assoc. 108(503):914–928. doi:10.1080/01621459.2013.787933.

Rényi A, Makkai-Bencsáth Z. 1984. A diary on information theory. Budapest: Akadémiai Kiadó. [accessed 2022 Feb 8]. https://academic.oup.com/blms/article-pdf/doi/10.1112/blms/20.4.380b/801811/20-4-380b.pdf.

Resnick S. 2007. Heavy-Tail Phenomena: Probabilistic and Statistical Modeling. New York, NY: Springer (Springer Series in Operations Research and Financial Engineering). [accessed 2020 Oct 24]. http://link.springer.com/10.1007/978-0-387-45024-7.

Selvaraj N, Mendelson Y, Shelley KH, Silverman DG, Chon KH. 2011. Statistical approach for the detection of motion/noise artifacts in Photoplethysmogram. In: 2011 Annual International Conference of the IEEE Engineering in Medicine and Biology Society. IEEE. p. 4972–4975. [accessed 2020 Nov 26]. http://ieeexplore.ieee.org/document/6091232/.

Shalizi CR. 2007. Maximum likelihood estimation for q-exponential (Tsallis) distributions. Arxiv Prepr math/0701854.

Sornette D. 1998. Multiplicative processes and power laws. American Physical Society. [accessed 2020 Dec 9]. https://arxiv.org/pdf/cond-mat/9708231.pdf.

Standish RK. 2008. Concept and Definition of Complexity. In: Yang A, editor. Intelligent Complex Adaptive Systems. IGI Global. p. 105–124. [accessed 2022 Feb 19]. https://www.igi-global.com/chapter/concept-definition-complexity/24185.

Stoev SA, Michailidis G, Taqqu MS. 2011. Estimating Heavy-Tail Exponents Through Max Self–Similarity. IEEE Trans Inf Theory. 57(3):1615–1636. doi:10.1109/TIT.2010.2103751. [accessed 2020 Nov 25]. http://ieeexplore.ieee.org/document/5714272/.

Tirnakli U, Borges EP. 2015. The standard map: From Boltzmann-Gibbs statistics to Tsallis

statistics. Sci Rep. 6:23644. doi:10.1038/srep23644. [accessed 2015 Oct 12]. http://arxiv.org/abs/1501.02459.

Tirnakli U, Tsallis C. 2020. Extensive Numerical Results for Integrable Case of Standard Map. Nonlinear Phenom Complex Syst. 23(2):149–152. doi:10.33581/1561-4085-2020-23-2-149-152. [accessed 2021 Dec 29]. https://doi.org/10.33581/1561-4085-2020-23-2-149-152.

Tsallis C. 2006. On the extensivity of the entropy Sq, the q-generalized central limit theorem and the q-triplet. Prog Theor Phys Suppl. 162:1–9. doi:10.1143/PTPS.162.1/5228515/162-1.PDF. [accessed 2022 Feb 6]. https://academic.oup.com/ptps/article/doi/10.1143/PTPS.162.1/1848078.

Tsallis C. 2009. Introduction to nonextensive statistical mechanics: Approaching a complex world. Springer Science & Business Media. [accessed 2019 May 14]. https://books.google.com/books?hl=en&lr=&id=qNIGnzcEiPMC&oi=fnd&pg=PA3&dq=Introduction+to+Nonextensive+Statistical+Mechanics&ots=BjyKOroF5K&sig=0kQqdmJVt5xbBT4uQ7-WmO9ZRlk.

Tsallis C, Plastino AR, Alvarez-Estrada RF. 2009. Escort mean values and the characterization of power-law-decaying probability densities. J Math Phys. 50(4):43303. doi:10.1063/1.3104063. [accessed 2019 Mar 27]. http://aip.scitation.org/doi/10.1063/1.3104063.

Vilela ALM, Wang C, Nelson KP, Stanley HE. 2019. Majority-vote model for financial markets. Phys A Stat Mech its Appl. 515:762–770. doi:10.1016/j.physa.2018.10.007. [accessed 2019 Nov 6]. https://www.sciencedirect.com/science/article/pii/S0378437118313451.

Viswanathan GM, Afanasyev V, Buldyrev S V., Murphy EJ, Prince PA, Stanley HE. 1996. Lévy flight search patterns of wandering albatrosses. Nature. 381(6581):413–415. doi:10.1038/381413a0. http://www.nature.com/doifinder/10.1038/381413a0.

Wilk G, Włodarczyk Z. 2000. Interpretation of the Nonextensivity Parameter q in Some Applications of Tsallis Statistics and Lévy Distributions. Phys Rev Lett. 84(13):2770–2773. doi:10.1103/PhysRevLett.84.2770. [accessed 2015 Sep 15]. https://link.aps.org/doi/10.1103/PhysRevLett.84.2770.

Zubillaga BJ, Vilela ALM, Wang C, Nelson KP, Stanley HE. 2022. A three-state opinion formation model for financial markets. Phys A Stat Mech its Appl. 588:126527. doi:10.1016/J.PHYSA.2021.126527.